\documentstyle[12pt]{article}
\begin{document}


\def\gw{$\gamma\gamma\rightarrow W^+W^-$}
\def\gwl{$\gamma\gamma\rightarrow W^+_{L} W^-_{L}$}
\def\gz{$\gamma\gamma\rightarrow ZZ$}
\def\gzl{$\gamma\gamma\rightarrow Z_L Z_L$}
\def\gpi{$\gamma\gamma\rightarrow \pi\pi$}
\def\M{\cal M}

\begin{titlepage}
\title{\bf The reactions \gwl\ and \gzl\ in $SU(N)$ strongly interacting
theories}
\author{
{\bf John F. Donoghue}\thanks{e--mail: donoghue@phast.umass.edu}\ \ \  {\bf
and Tibor Torma}\thanks{e--mail: kakukk@phast.umass.edu} \\[2mm]
Department of Physics\\University of Massachusetts\\Amherst,\ \ \ MA 01003
\ \ U.S.A.}
\date{}
\maketitle
\def\baselinestretch{1.15}
\begin{abstract}
Building on recent phenomenology of \gpi, we discuss the expectation for two
photon production of longitudinal gauge boson pairs in $SU(N)$ technicolor
theories. The treatment involves a matching of dispersive techniques with the
methodology of chiral perturbation theory.
\end{abstract}
\vskip-15.5cm
\rightline{ UMHEP-398}
\rightline{ September 1993}
\vskip3in
\end{titlepage}
\newpage
\def\baselinestretch{1}
\section*{\bf I.~Introduction}
\footnotetext{This work has been supported in part by the National Science
Foundation}
If the symmetry breaking part of the Standard Model is strongly interacting,
it becomes important to study reactions involving longitudinal gauge bosons.
These reactions may yield information on the underlying mechanism of symmetry
breaking. The case of $W^+_{L} W^-_{L} \rightarrow W^+_{L} W^-_{L}$ scattering
has been extensively explored~\cite{int1}. Photon production of longitudinal
bosons, \(\gamma\gamma\rightarrow W_{L} W_{L}\), may also be an interesting
reaction to study, and it has been discussed increasingly
recently~\cite{int3},\cite{her1}.
Because the equivalence theorem~\cite{eq1}\cite{eq2}\cite{eq3}\cite{eq4}
relates the longitudinal gauge boson coupling to those of the original scalars
/pseudoscalars of the underlying theory when the energy is high enough, the
discussion may also be phrased in terms of the production of the Goldstone
bosons. The corresponding reactions in QCD, \gpi, has recently had extensive
theoretical investigations as well as comparison with experiment, and the
connection between the chiral constraints and dispersion theory have been
resolved. Since we appear to have now a theoretical control over this reaction,
it is of interest to display the expectations for an $SU(N)$ QCD-like theory,
scaled up to $TeV$ energies. If the symmetry breaking sector is due to fermion
condensates, such as may occur in technicolor, these expectations could
possibly be realized by Nature. If not, these results may still form a useful
contrast with other possibilities, such as a strongly interacting Higgs theory.
In this paper we apply the formulation of Ref.~\cite{2} to the reactions
$\gamma\gamma\rightarrow WW$.

There have been two recent papers~\cite{int3},\cite{her1} which show several
similarities with our work, with the most extensive exploration of the
phenomenological consequences being given by Herrero and
Ruiz-Morales~\cite{her1}. Our paper is different in its dispersive treatment
and in its use of recent experimental and theoretical work on the \gpi\
reaction. We also emphasize the importance of techni-$f_2$-like resonances
in the direct channel.

The paper proceeds as follows. In Sect.~II., we review the essential features
of the QCD reaction \gpi, and explain in general how this may be transformed
to an $SU(N)$ theory at $TeV$ scales. Sect.~III. describes the details of the
calculation, and presents the results. In an appendix we discuss the use of the
equivalence theorem in loop diagrams.

\section*{\bf II.~\gpi\ and a model for \(\gamma\gamma\rightarrow WW,ZZ\)}

The two important tools in the analysis of \gpi\ are chiral symmetry and
dispersion relations. Early work describing the one loop chiral calculation of
\(\gamma\gamma\rightarrow \pi^0\pi^0\) yielded results which, while valid for
small enough pion mass, receive significant corrections even at the physical
threshold. These corrections have been seen in a two loop chiral
calculation~\cite{twoloop} and in dispersive studies~\cite{2},\cite{7}, and are
also needed experimentally. Since the chiral results for this particular
reaction are modified at such a low energy, one needs to use other methods in
order to work at higher energies. Dispersion relations are useful in this
regard. In fact, the two techniques work well together, since chiral symmetry
allows one to identify the substraction constants which appear in dispersive
treatments.

However, because dispersion relations are a methodology, not an underlying
theory, the result are only as good as the inputs to the calculations. While
theorists have become reasonably adept at identifying the appropriate input,
it is valuable to have experimental data to verify the results.

The starting point is a dispersive representation for the \gpi\ S-wave
amplitude with isospin~$I$,
{\begin{equation}
f_I(s)=t_I(s)+D_I^{-1}(s)\{(c_I+d_I s)-\frac{s^2}{\pi}\int_{4m_\pi^2}^{
\infty}\frac{dx}{x^2}\frac{t_I(x)ImD_I(x)}{x-s-i\epsilon}\}
\label{mod1}\end{equation}}
originally given by Morgan and Pennington~\cite{5}. Here, $t_I(s)$ is a real
function which contains all of the correct cuts of the t-channel process
$\gamma\pi\rightarrow\gamma\pi$, $D_I^{-1}(s)$ is the Omn\`es function which
can be obtained from $\pi\pi$ scattering, and $c_I$, $d_I$ are substraction
constants. The dispersion relation is valid below inelastic thresholds. In
order to have a succesful calculation, one must specify correctly the three
ingredients listed above. In Ref.~\cite{2}, this was accomplished as follows.
By comparison with the one loop calculation in chiral perturbation theory, the
substraction constants are identified as
\begin{eqnarray}
c_0 = c_2 = 0 \\
d_0 = d_2 = \frac{2}{F^2_\pi} (L^r_9+L^r_{10}) +
\frac{t^{CA}_I(0)}{12\pi m^2_\pi}
\label{in1}\end{eqnarray}
Here, $L^{r}_9$ and $L^{r}_{10}$ are coefficients in the most general $O(E^4)$
chiral Lagrangian, renormalized at scale $\mu$. The combination presented in
$d_0,d_2$ is invariant under a change in $\mu$. Eqn.~(\ref{in1}) is valid up to
higher order chiral correction of order $m^2_\pi$. The Omn\`es function for
$I=0$ was obtained by Gasser~et~al.~\cite{gas14} from a detailed fit to the
experimental data on $\pi\pi$ scattering, also incorporating the correct
chiral behavior for the $\pi\pi$ amplitudes at low energy. The $I=2$ Omn\`es
function is not very important in the calculations and was obtained by a simple
Pad\'e approximation fit to the low energy $\pi\pi$ data. The remaining
important ingredient, $t_I(s)$, is required by Low's theorem to approach the
Born amplitude at low energy. At higher energies other intermediate states,
due to $\rho, \omega$ exchange (i.e.~$\gamma\pi\rightarrow\rho,
\rho\rightarrow\gamma\pi$) are known to be present and are included with the
coupling constant determined from \(\rho,\omega\rightarrow\gamma\pi\) data.
The combination of these ingredients is sufficient to accurately describe the
data up to $s=1 \, GeV$ (see Fig.~1). For the region from 1~to~$1.5 \, GeV$ the
dominant feature is the $f_2(1270)$ resonance in the $\pi\pi$ D-wave.

We wish to scale this result up in energy in order to describe
$\gamma\gamma\rightarrow WW$ in one possible realization of a strongly
interacting symmetry breaking sector. Specifically, our target theory is an
$SU(N)$ gauge theory with a single doublet of fermions. Such a theory possesses
an $SU(2)_L\times SU(2)_R$ global chiral symmetry, where dynamical symmetry
breaking will also break the $SU(2)_L$ symmetry of the Standard Model. The
pions of this gauge theory (called technipions below) are the Goldstone bosons
which form the longitudinal components of the $W^\pm, Z^0$. This theory is the
prototype of technicolor theories. As is well known, more ingredients must be
added if one is to attempt to understand fermion masses. However, at present
there is no simple way to extend the theory in order to obtain all desired
results. The most essential ingredient for our calculations is the particle
spectrum, specifically the techni-$\rho$, techni-$\omega$ and techni-$f_2$,
and the absence of light states ($L$) which the technipions would scatter
inelastically, $\pi_T\pi_T\rightarrow LL$. If extensions to the one doublet
technicolor do not strongly modify these features, our calculation can remain
valid.

The equivalence theorem states that the scattering amplitudes of longitudinal
gauge bosons are related at high energy to those of the Goldstone bosons of the
theory, in this case, to the technipions:
{\begin{equation}
{\M}(\gamma\gamma\rightarrow W^i_LW^j_L) = {\M}(\gamma\gamma\rightarrow
\pi^i_T\pi^j_T) + O(\frac{M_W}{\sqrt{s}})
\end{equation}}

The restriction to high energies implies that any dependence on $M_W$ must be
small, and we will assume this to be true at $\sqrt{s}\geq 400 \, GeV$. We will
also see that our ability to make predictions fail at about $\sqrt{s}=3 \,
TeV$. Thus, we may expect that our calculations will be best applied between
these two extremes.

In order to obtain the $\gamma\gamma\rightarrow WW$ amplitude we must i)~scale
energy variables up to the appropriate $TeV$ scale and ii)~modify the three
ingredients described in regards the \gpi\ calculations. The former is the
simplest and best known. The vacuum expectation value \(v=246 \, GeV\) plays
the same role in the Standard Model as \(F_\pi=92.3 \, MeV\) plays in QCD.
However, in the large-$N_c$ counting rules of QCD, \(F_\pi=c\Lambda_{QCD}
\sqrt{N_{TC}}\), where $c$ is a dimensionless constant and \(\Lambda_{QCD}\)
is the energy scale of QCD. Scaling this would require
\(v=c\Lambda_{TC}\sqrt{N_{TC}}\) with the same constant $c$ and with
\(\Lambda_{TC}\) being the energy scale of the
\(SU(N_{TC})\) theory. The scaling for all the normal resonances
(\( m_{X_T}\sim\Lambda_{TC}\)) is then
{\begin{equation}
m_{X_T} = m_X\frac{\Lambda_{TC}}{\Lambda_{QCD}} = m_X\frac{v}{F_\pi}\sqrt{
\frac{N_c}{N_{TC}}}
\end{equation}}
This is the scaling for \(X=\rho, \omega, f_2\).

The pion mass is of course an exception, since the quark masses of QCD have no
counterpart in the technicolor theory. The technipion would be massless, but
become the longitudinal component of the W. For kinematic reasons we will use
\(m_{\pi_T}=M_W\). However, the difference between \(m_{\pi_T}=0\) and
\(m_{\pi_T}=M_W\) is an \(O(\frac{M_W}{\sqrt{E}})\) correction which is anyway
below those predicted by the equivalence theorem.

In order to provide the remaining ingredients for the amplitudes (i.e.~the
substraction constant $d_I$, the Omn\`es function \(D^{-1}_I(s)\) and the
elementary amplitude \(t_I(s)\) ), two sets of corrections must be made
besides simply scaling the energy: i)~Even for \(N_{TC}=N_c=3\), we must
correct for the change in the technipion mass since
{\begin{equation}
\frac{m_{\pi_T}}{\Lambda_{TC}}\ll\frac{m_\pi}{\Lambda_{QCD}}
\end{equation}}
and ii)~in addition, if \(N_{TC}\neq N_c\) we must correct for the changes in
the scattering amplitude. The details and the results of each of the changes
is given in Sect.~III., while below we describe the general methods that we
employ.

In some portions of the scattering amplitude it is easy to make the
substitutions \((m_\pi,F_\pi)\Rightarrow(m_{\pi_T}=m_W,v)\). In particular we
have analytic expressions for the Born and resonance exchange amplitudes and
therefore these present no problem in rescaling the Goldstone boson mass (we
assume that the coupling constants for \(\rho,\omega\rightarrow\pi\gamma\)
have no significant dependence on the pion mass). However, the Omn\`es function
is more difficult because it is based on physical data. As we explain in more
detail in the next section, we get around this by using chiral symmetry to
describe the pion mass dependence of the Omn\`es function at very low energy,
and smoothly matching on the experimental data at high energy (where the mass
dependence should be unimportant).

The substraction constants also depend on the pion mass, although in an
indirect way. The renormalization prescription, adopted in Ref.~\cite{ren},
creates an implicit dependence in the renormalized chiral coefficients
\(L_j^r\) on \(\ln m_\pi^2\). This is easily corrected for via
{\begin{equation}
L^r_j(m_2) = L^r_j(m_1) + \frac{\Gamma_j}{16\pi} \ln\frac{m_2}{m_1}.
\end{equation}}

The other set of corrections involve accounting for \(N_{TC}\neq 3\). For this
we use the large-N counting rules wherever possible. Most of these are obvious.
The case that requires the most thought is the $I=0$ Omn\`es function as it is
determined by experimental data without a clear resonant behaviour. Our
procedure here was to use chiral symmetry plus a Pad\'e fit to the experimental
Omn\`es functions in order to shift $N_{TC}$. The low energy chiral behavior is
independent of $N_{TC}$, since $v$ is held fixed but the first chiral
correction is linear in $N_{TC}$. We note that our use of the Pad\'e
approximation only as a guide for estimating corrections to the experimental
Omn\`es functions.

The above describes the main ingredients of our model. Basically we try to
normalize our method to the succesful treatment of \gpi, and modify this by
well-defined corrections. We implement this in the next section.

\section*{\bf III.~Details of the calculation}

\subsection*{- {\it basics}}

In this section we describe in detail how we calculate the \gw \ and \gz \
cross sections. We use the notation of Fig.~2 and we find that the gauge
determined by
{\begin{equation}
k_1\cdot \epsilon_2 = k_2\cdot \epsilon_1 = 0
\label{1}\end{equation}}
greatly simplifies our calculations. The scattering amplitude with totally
polarized incoming photons~\cite{3}, is required by Bose symmetry, Lorentz
and gauge invariance to be equal to
$\epsilon_{\mu}^1\epsilon_{\nu}^2\M^{\mu\nu}$ with

{\begin{eqnarray}
\M^{\mu\nu} & = & 4 i e^2 \{ A[k_2^\mu k_1^\nu-(k_1k_2)g^{\mu\nu}] + C
\epsilon^{\mu\nu\alpha\beta}k_{1\alpha}k_{2\beta}\nonumber\\
 &   &+ B[\frac{(p_1k_1)(p_1k_2)}{(k_1k_2)} +p_1^\mu p_1^\nu -
\frac{(p_1k_1)}{(k_1k_2)} k_2^\mu p_1^\nu -\frac{(p_1k_2)}{(k_1k_2)}k_1^\nu
p_2^\mu]\}
\label{2}\end{eqnarray}}

In the neutral case an additional symmetrization in pion momenta is required.
Parity invariance of EM and strong interactions requires $C=0$ and the parity
violation of the weak interactions is not important for the production of
longitudinal gauge bosons. The polarization vectors of a totally polarized
incoming photon are described in the helicity basis by two complex numbers as
{\begin{eqnarray}
\epsilon_1^\mu & = & e_1^{(-)}\bar{\epsilon}^\mu
+ e_1^{(+)}\epsilon^\mu\nonumber\\
\epsilon_2^\mu & = & e_2^{(-)}\epsilon^\mu +
e_2^{(+)}\bar{\epsilon}^\mu\label{3}
\end{eqnarray}
Using this form and trivial kinematical considerations we arrive at
{\begin{eqnarray}
\M & = & 2ie^2 \{ (As-m_\pi^2B)[e_1^{(+)}e_2^{(+)}+e_1^{(-)}e_2^{(-)}]+
\nonumber\\
&   & +Bp^2\sin^2\theta[e_1^{(+)}e_2^{(-)}e^{2i\phi}+e_1^{(-)}e_2^{(+)}e^{-2i
\phi}]\}
\label{4}\end{eqnarray}}
where the helicity conserving amplitude, $As - m_\pi^2 B$, and the helicity
flipping $B p^2$ terms are clearly separated. It is now
straightforward to calculate the cross sections~\cite{2}
{\begin{equation}
\frac{d\sigma}{dt} = \frac{S\beta}{16(2\pi)^2s} {|\M|}^{2}
\label{5}\end{equation}}
where the statistical factor $S=1$ for the charged pion final state and
$S=1/2$ for the neutral case. As dedicated \(\gamma\gamma\) colliders mostly
do not yield strongly polarized photons~\cite{tel15}, we neglect photon
polarization effects. Then, for completely unpolarized and uncorrelated
incoming photons one has~\cite{2}
{\begin{equation}
(\frac{d\sigma}{dt})_{unpol.} = \frac{2\pi S\alpha^2}{s^2}\{|As-m_\pi^2B|^2
+\frac{|B|^2}{s^2}(m_\pi^2-tu)^2 \}
\label{6}\end{equation}}

As we have spin-1 incoming particles we must use the full $lm$ partial wave
expansion. The only existing waves are with $m=2,0,-2$ and we choose the
normalization of the partial amplitude as
{\begin{eqnarray}
\M & = & \sqrt{4\pi}2ie^{2} \sum_{l}\{Y_{l0}(\theta,\phi)[e_1^{(+)}e_2^{(+)}+
e_1^{(-)}e_2^{(-)}] f^{l0}(s) +\nonumber\\
&   & + Y_{l2}(\theta,\phi) [e_1^{(-)}e_2^{(+)}] f^{l2}(s)+
Y_{l,-2}(\theta,\phi) [e_1^{(+)}e_2^{(-)}] f^{l,-2}(s)\}
\label{7}\end{eqnarray}}

Inverting this formula we may connect the partial wave amplitudes to $A$ and
$B$:
{\begin{eqnarray}
f^{l0}&=&\frac{\sqrt{2l+1}}{2}\int_{-1}^{1}dz P_l(z)(As-m_\pi^2B)\nonumber\\
f^{l2}(s)&=&\sqrt{\frac{2l+1}{(l-1)l(l+1)(l+2)}}\frac{1}{2}
\int_{-1}^{1}dz\frac{1-z^2}{2}P_l^2(z)B
\label{8}\end{eqnarray}}

Using isospin symmetry we may build the isospin-0 and isospin-2 amplitudes
(there are no~I=1 waves)
{\begin{eqnarray}
f_C^{lm}(s) & = &  \frac{2}{3}f_0^{lm}(s)+\frac{1}{3}f_2^{lm}(s)\nonumber\\
f_N^{lm}(s) & = & -\frac{2}{3}f_0^{lm}(s)+\frac{2}{3}f_2^{lm}(s)
\label{9}\end{eqnarray}}

The idea of dealing with unknown technicolor models is to use their similarity
to QCD pion production. There, according to the findings of~\cite{2}, the main
ingredients in the $.5$--$1 \, GeV$ region (which will correspond to the $TeV$
range in our case) are the Born+seagull amplitudes (only present in the charged
channel), the vector and axial vector exchanges as dictated by a vector
dominance model and final state pion rescattering which gives an important
contribution to the neutral channel even near threshold. This model is known
to reproduce the chiral results for low energy and assumes that rescattering
is only important in the S-wave. For this reason we neglect all higher-wave
rescattering and in the S-wave we also assume the validity of Watson theorem.
In QCD this is true even above multipion thresholds and only breaks down at
two-kaon threshold. As in our model we have no kaon we may suppose its validity
all the way up to the $f_2$ mass range.

Then, in a way similar to~\cite{2}, we add the Born and full vector-dominance
amplitudes to S-wave rescattering. The latter is handled by writing a doubly
subtracted dispersion relation for $f_I(s)\equiv f_I^{00}$~\cite{2}

{\begin{equation}
f_I(s)=p_I(s)+D_I^{-1}(s)\{(c_I+sd_I)-\frac{s^2}{\pi}\int_{4m_\pi^2}^{\infty}
\frac{dx}{x^2}\frac{p_I(x)ImD_I(x)}{x-s-i\epsilon}\}
\label{10}\end{equation}}
where $p_I(s)$ is the partial amplitude from Born~+~vector dominance
contributions; the Omn\`es functions
{\begin{equation}D_I(s)=exp\{-\frac{s}{\pi}\int_{4m_\pi^2}^{\infty}
\frac{ds^{'}}{s^{'}} \frac{\delta_I(s)}{s^{'}-s-i\epsilon}\}
\label{11}\end{equation}}
establish a connection to the $\pi\pi$ elastic phase shifts. The question of
subtraction constants is reviewed in~\cite{2} and we simply quote the results,
\(c_I\equiv 0\) (to satisfy Low's theorem) and
{\begin{equation}
d_I = \frac{t_I^{CA}(0)}{12m_\pi^2}
\label{12}\end{equation}}
to satisfy the chiral constraint (\(t_I^{CA}\) are the Weinberg scattering
amplitudes so $d_I$ doesn't really depend on~$m_\pi$). Note that since
vector and axial vector mesons generate \(L_9+L_{10}\), the effect of the
chiral coefficients \(L_9+L_{10}\) also appears in $p_I(s)$ rather than in
$d_I$, in contrast to the form in Eqn.~2.

The particular form of the vector dominance model we use is also taken
from~\cite{2}, together with the QCD vector meson--pion--photon coupling
constants $R_\omega=1.35 \, GeV^{-2}$ and $R_\rho=0.12 \, GeV^{-2}$. It
includes the $\omega$ and $\rho$ mesons in narrow-width approximation and also
\(a_1(1270)\) which is found to have a large contribution to
$\gamma\gamma\rightarrow \pi^0\pi^0$. Their contribution into $A$ and $B$ are
worked out in full detail in~\cite{2}, Eqns.~(35),(37):
{\begin{eqnarray}
sA^0 & = & -\frac{s}{2} \sum_{V=\rho,\omega} R_V (\frac{m_\pi^2+t}{t-m_V^2}+
\frac {m_\pi^2+u}{u-m_V^2})\nonumber\\
B^0 & = & -\frac{s}{2} \sum_{V=\rho,\omega} R_V (\frac{1}{t-m_V^2}+
\frac{1}{u-m_V^2})
\label{13}\end{eqnarray}}
and
{\begin{eqnarray}
sA^+ & = & -\frac{s}{2} R_\rho (\frac{m_\pi^2+t}{t-m_\rho^2} +
\frac {m_\pi^2+u}{u-m_\rho^2}) + m_A^2 s \frac{L_9^r+L_{10}^r}{F_\pi^2}
(\frac{1-\frac{m_\pi^2+t}{2m_A^2}}{t-m_A^2} + \frac{1-\frac{m_\pi^2+u}
{2m_A^2}}{u-m_A^2})\nonumber\\
B^+ & = & - (\frac{1}{t-m_\pi^2} + \frac {1}{u-m_\pi^2}) -
\frac{sR_\omega}{2} (\frac{1}{t-m_\rho^2} + \frac {1}{u-m_\rho^2})
\label{14}\end{eqnarray}}

We cut off the cross section is at \(Z\equiv cos\Theta=0.6\). This, in
addition to trying to mimic the experimental limitation, serves another
purpose: most of the $W^+W^-$ events go into forward direction and constitute
a huge background which is greatly reduced this way~\cite{11}. The cross
section is given by
{\begin{equation}
\int_{t_a}^{t_b}dt \frac{d\sigma}{dt} \ \ {\rm with} \ \ t_{b,a}=m_\pi^2-
\frac{s}{2}(1\mp \beta Z)
\label{15}\end{equation}}

This integration can be made by hand, and we calculated the dispersive
integral numerically.

\subsection*{- {\it 1/N: meson masses and couplings}}

As we mentioned in Sect.~II., QCD results have to be generalized to different
numbers of colors, and we take the required additional information from the
leading $1/N_{TC}$ behaviour. In the following we find the leading $1/N_{TC}$
behaviour of the couplings of vector and axial vector mesons, namely $R_\rho$,
$R_\omega$ and $L_9+L_{10}$. They are restricted by the Weinberg sum
rules~\cite{4}, which in the narrow width approximation read
{\begin{equation}
F_\pi^2 = F_\rho^2 - F_{a_1}^2 \ \ and \ \ F_\rho^2 m_\rho^2 = F_{a_1}^2
m_{a_1}^2
\label{16}\end{equation}}

These restrictions imply that $F_V$ and $F_A$ are constants in $1/N_{TC}$.
Their connection to the constants of equations (\ref{13}) and (\ref{14}) can
be seen simply on dimensional grounds,
{\begin{equation}
R_V = c \frac{F_V^2}{m_V^4}
\label{17}\end{equation}}
with $c$ being a dimensionless constant for vectors and $F_A$ can be taken
from~\cite{4}

{\begin{equation}
L_9+L_{10} = \frac{F_A^2}{m_A^2}
\label{18}\end{equation}}

All this means that the needed $1/N_{TC}$ rules are
{\begin{equation}
L_9+L_{10} \sim N_{TC} \ \ and \ \ R_V \sim \frac{N_{TC}}{m_\rho^2}
\label{19}\end{equation}}

\subsection*{- {\it pion scattering amplitudes}}

There is an important and interesting physics in the outgoing pion
rescattering process. To describe it, we need to know the Omn\`es functions
from somewhere. These are connected to the $\pi\pi$ elastic phase shifts as
long as the elastic process dominates. Our description is consequently limited
to this energy region; fortunately this means \(\sim~1 \, GeV\) in QCD. In
terms of technicolor this scales up to $2.6 \, TeV$ for $N_{TC}=3$, even for
$N_{TC}=10$ we still may expect a relatively good description at $1.4 \, TeV$.
The fact that in our model there are no kaons may push that limit even a
little further. Nevertheless, we do not expect to have the right description
above the (techni)$f_2$ resonance and we simply neglected the tail of the
dispersion integral which comes from energies above $3 \, TeV$. Several reruns
with different energy cutoffs showed little effect.

For the $N_{TC}=3$ case a simple rescaling of the known Omn\`es functions is
possible. However, if we want to generalize our statements for higher
$N_{TC}$'s
--~there is no reason to expect that technicolor is just a copy of QCD~-- we
need a deeper understanding of the phase shifts. For the region
\(s<0.1 \, GeV^{-2}\) (as in QCD) chiral perturbation theory gives a
satisfactory description, but a second order (in $s$) calculation shows that
some uniterization is necessary. We did this, using [1,1] Pad\'e
approximation~\cite{2}
{\begin{eqnarray}
Re D_I^{Pad\acute{e}}(s) & = & 1 -sc_I + t_I^{CA}(s)\frac{2}{\pi}
[\frac{\beta}{2}\ln{\frac{s(1+\beta)^2}{4m_\pi^2}}-1]\nonumber\\
Im D_I^{Pad\acute{e}}(s) & = & -\beta t_I^{CA}(s)
\label{20}\end{eqnarray}

Although we know that this procedure only gives an approximate result, we use
it as a vehicle for introducing modifications to the experimental Omn\`es
functions. In (\ref{20}), the constants $c_I$ are expressed by a certain linear
combinations of chiral coefficients
{\begin{equation}
c_I=\frac{l_I^r(m_\pi)}{f_\pi^2}
\label{21}\end{equation}}
and they pick up logarithms in the chiral limit.

In the standard chiral renormalization scheme, the chiral coefficients
\(l_I\) contain factors of \(\ln m^2_\pi\). For our purposes it is better to
remove these so that we define  \(\bar{c}_I=\frac{\bar{l_I^r}}{f_\pi^2}\), with
{\begin{eqnarray}
Re D_I^{Pad\acute{e}}(s) & = & 1 -s\bar{c}_I(\mu) + t_I^{CA}(s)\frac{2}{\pi}
[\frac{\beta}{2}\ln{\frac{s(1+\beta)^2}{4\mu^2}}-1] +\nonumber\\
                   &   & [\frac{\beta}{\pi}t_I^{CA}(s) - \frac{\gamma_I}
{32\pi^2f_\pi^2}s] \ln{\frac{\mu^2}{m_\pi^2}}
\label{22}\end{eqnarray}}
where the numbers $\gamma_I$ govern the chiral logarithms
{\begin{equation}
\bar{l_I^r} = l_I^r + \frac{\gamma_I}{32\pi^2}\ln{\frac{\mu^2}{m_\pi^2}}
\label{23}\end{equation}}

As the Omn\`es functions are known to have a finite chiral limit (i.e.
\(\delta_I(s)=s\times({\it finite})\)) we conclude that the quantity in the
square brackets vanishes in that limit, yielding $\gamma_0=2$ and
$\gamma_2=-1$.

The chiral amplitude also depends directly on $N_{TC}$. To determine this
dependence we use the succesful model of the chiral constants, expressed in
terms of resonance saturation. This again comes from matching chiral symmetry
with vector dominance~\cite{12}
{\begin{equation}
l_I^r(m_\rho) = (const) \frac{\Gamma_\rho F_\pi^4}{m_\rho^5} = O(N_c)
\label{24}\end{equation}}

This form requires that the leading $1/N_{TC}$ behaviour to be
{\begin{equation}
\bar{c}_I^{TC}(\mu) = \frac{N_{TC}}{3} \{ {\frac{f_\pi^2}{F_\pi^2}
\bar{c}_I^{QCD}(\mu^{QCD}) + \frac{\gamma_I}{16\pi^2F_\pi^2}
\ln(\frac{\mu^{QCD}}{m_\rho^{QCD}}\frac{m_\rho^{TC}}{\mu})} \}
\label{25}\end{equation}}

As an input we still need a value for \(\bar{c}_I^{QCD}(\mu^{QCD}=140 \,
MeV)\).
Plotting several values on Fig.~9, using the method of trial and error, we have
the best fit when \(\bar{c}_0^{QCD}(\mu^{QCD}=140 \, MeV)=1.5\pm0.2\) and
\(\bar{c}_2^{QCD}(\mu^{QCD}=140 \, MeV)=-1.7\pm0.5\).

\subsection*{- {\it Inclusion of $f_2$}}

In \(\gamma\gamma\rightarrow\pi\pi\) scattering an important feature in the
direct channel is the $f_2$ resonance, so we may reasonably suspect that
incorporating a corresponding technimeson may be important. This is especially
true for larger $N_{TC}$'s where meson resonances come down in mass and become
narrower.

As $f_2$ is a tensor particle its couplings require special consideration. It
is not present in the crossed channels as C-invariance forbids
\(f_2\rightarrow\gamma\pi\), and, as we saw that
\({\cal F}(\gamma\gamma\rightarrow\pi\pi\)) is generally described in terms of
two amplitudes, the description of $f_2$ requires information on two coupling
constants, in terms of the two possible photon helicity differences. When we
turn to the existing experimental data on
\(\gamma\gamma\rightarrow\pi\pi\)~\cite{10}, we see that there is a controversy
in determining the ratio of different waves near $f_2$ mass. It had been
generally accepted that the interaction is pure \(|JJ_3>=|22>\), and
although~\cite{10} argues that the presence of a \(|20>\) interaction is
favored, but its significance is rather poor \((\chi^2/N_{DF}\sim1.8-2.0\)
for \(N_{DF}=177)\) and as long as data with scattering angle cut \(|Z|<0.6\)
are concerned, they may be explained by the conservative hypothesis of $|22>$
dominance. This assumption means, in terms of the amplitudes of (\ref{6}), that
we have no impact of $f_2$ in \((As-m_\pi^2B)\) and the {\it same}
$t$-independent quantity \({\cal F}_{22}(s)\) is added to $B_C$ and $B_N$. For
simplicity we suppose that this amplitude is of a pure Breit-Wigner form
{\begin{equation}
{\cal F}_{22}(s) = \frac{{\cal F}_{22}}{s-(m_{f_2}^2-\frac{i}{2}
\Gamma_{f_2})^2}
\label{33}\end{equation}}.

In the case of a single-channel resonance ${\cal F}_{22}$ is required to be
real. Its value is determined by the well-known Breit-Wigner formula
{\begin{equation}
\sigma_{\pi^+\pi^-}^{unpol} =\frac{40\pi}{s}
\frac{\Gamma_{f_2\rightarrow\pi\pi}\Gamma_{f_2\rightarrow\gamma\gamma}
m_{f_2}}{|s-m_{f_2}^2+im_{f_2}\Gamma_{f_2}|^2}
\label{32}\end{equation}}
and this is to be compared to the {\it pure} Breit-Wigner form of the amplitude
{\begin{equation}
\sigma_{\pi^+\pi^-}^{unpol} = \frac{2\pi\alpha^2}{s^2}
\frac{|{\cal F}_{22}|^2}{|s-m_{f_2}^2+im_{f_2}\Gamma_{f_2}|^2}
\int_{t_a}^{t_b}dt \frac{(m_\pi^4-tu)^2}{s^2}
\label{26}\end{equation}}
This implies
{\begin{equation}
|{\cal F}_{22}| = \frac{20}{\alpha m_{f_2}} \sqrt{
\Gamma_{f_2\rightarrow\pi\pi} \Gamma_{f_2\rightarrow\gamma\gamma} }
\label{27}\end{equation}}
and the sign of ${\cal F}_{22}$ is found by fitting the results to the
experimental data. To this end, we plotted on Fig.~3 and Fig.~4 the full
\(\gamma\gamma\rightarrow\pi\pi\) cross sections with both negative and
positive ${\cal F}_{22}$. Although the neutral pion channel is inconclusive,
comparing Fig.~3 to the corresponding experimental values (that is, to Fig.~4a
in~\cite{10}), we clearly see that \({\cal F}_{22}\) must be negative as the
vector meson contribution and the resonance have a clear destructive
interference just above, while a positive interference below resonance. The
dispersive integral gives very little contribution in this range. The
correctness of the shape of our curve and its numerical height supports our
assumption that we may neglect $|10>$ contributions to the s-channel $f_2$
contribution.

We take the numerical values as given by the Particle Data Group~\cite{13}
{\begin{equation}
\Gamma_{f_2\rightarrow\gamma\gamma}=2.6\, keV \ , \
\Gamma_{f_2\rightarrow\pi\pi}=0.84 \Gamma_{f_2} \ , \   \Gamma_{f_2}=185 \, MeV
\label{28}\end{equation}}

These considerations allow us to calculate the impact of a QCD-like
techni-$f_2$ on \(W_L^+W_L^-,ZZ\) production. We still have to see how
${\cal F}_{22}$ changes with $N_{TC}$. As we have no chiral prediction at
this high energy we go back to conventional $1/N$ arguments. Then, as seen
from Fig.~5,
{\begin{equation}
\Gamma_{f_2\rightarrow\gamma\gamma}/\Gamma_{f_2\rightarrow\pi\pi} = O(e^4 N^2)
\label{29}\end{equation}}
which yields, in terms of the coupling constant
{\begin{equation}
|{\cal F}_{22}^{TC}|=\frac{20}{\alpha}\sqrt{\frac{\Gamma_{\gamma\gamma}^{TC}
\Gamma_{\pi\pi}^{TC}}{(m_{f_2}^{TC})^2}}=
\frac{20}{\alpha}\sqrt{\frac{\Gamma_{\gamma\gamma}^{TC}\Gamma_{\pi\pi}^{TC}}
{(m_{f_2}^{QCD})^2}}=|{\cal F}_{22}^{QCD}|
\label{30}\end{equation}}

This enhancement of the \(\gamma\gamma\) branching ratio makes the techni-$f_2$
resonance {\it the} important feature of $ZZ$ production at higher $N_{TC}$.

\subsection*{- {\it Vector meson form factors}}

The vector meson dominance amplitude as given by~(\ref{13},\ref{14}) has an
incorrect high energy behaviour. For example, a simple counting of factors of
$s$ shows that \(\sigma_N\) grows as $O(s)$ and this obviously violates
unitarity. To cure this, we realize that all vector mesons are extended
objects of about the same size, and their interactions should be described by
form factors. The same power counting shows that a factor of \((m_\pi^2+t)\)
in the vector propagator numerators is responsible for the leading high energy
behaviour, and this goes back to the momentum-squared terms describing
longitudinal vector meson contributions. Fig.~8 shows what we get using
'pointlike' vector mesons (e.g.~constant form factors). The high energy
increase obviously contradicts to the experimental data.

For each of the vertices in Fig.~6 we have a separate form factor. There
action is described by taking the vector meson coupling constants $R_\omega$,
$R_\rho$ and $R_A$ as functions of $t$ in the form
{\begin{equation}
G=\frac{G(t=0)}{1-t/M^2}
\label{31}\end{equation}}

As we will see these form factors have little effect below \(s=m_{f_2}\), this
simple form is acceptable with \(M\sim m_\rho\). The integrations, similar to
the ones in (\ref{13}), yield formulas which, in some energy ranges, represent
the difference of large and almost cancelling quantities, so their evaluation
has been done by 16-byte arithmetics. Indeed, as curves on Fig.~7 shows, only
cross section near and above $m_{f_2}$ are affected.

\section*{\bf IV.~Discussion}

A high luminosity \(\gamma\gamma\) collider is not yet feasible at the energies
required for gauge boson production. However the concept is attracting
attention and it is possible that it may in the future become a reality. If a
Higgs boson is not found at lower energies, the study of
\(\gamma\gamma\rightarrow W^+W^-,Z^0Z^0\) will be one of the physics goals of
such a facility. The present calculation is the best description that we are
able to provide for these reactions in a particular type of theory, closely
related to QCD. Most of the features of the QCD reaction \gpi\ are visible in
our result, although they are somewhat modified by changes in the Goldstone
boson mass and the possibly different gauge group.

The main results of our calculations are shown on Fig.~10~a,b. Comparing them
to the analysis of experimental possibilities in Ref.~\cite{her1}, we conclude
that, with the presently envisaged parameters of a \(\gamma\gamma\) collider,
which may be realized on a linear $TeV$ electron accelerator with laser beam
backward scattering~\cite{tel15}, the \gwl\ process is clearly visible in two
regions. i)~below $\sim 500~\, GeV$, where the validity of our calculations is
restricted by the applicability limit of the equivalence theorem, and ii) on
the $f_2$ resonance. This is a huge resonance and its existence would be a
clear signal for a technicolor-like model, yielding in the same time
information on the value of $N_{TC}$. In the case of the \gzl\ reaction, a
similar statement says that the only really observable feature is the $f_2$
resonance, at presently proposed luminosities.

It would also be interesting to contrast these results with a different
strongly interacting symmetry breaking mechanism, that of a very heavy Higgs
boson. However, the latter theory may even be difficult to define, as the
studies of triviality of the $\Phi^4$ theory mean that a very heavy Higgs is
not possible unless other new physics enters the theory.

\newpage
\section*{\bf Appendix: loop diagrams and the equivalence theorem}

The equivalence theorem~\cite{eq1}\cite{eq2}\cite{eq3}\cite{eq4}} says
that the interactions of {\it external} longitudinal gauge bosons and the
corresponding Goldstone bosons are equal at high energy. However, to our
knowledge, the proofs do not address the issue of whether there is also an
equivalence for particles inside loop diagrams. For example, in chiral
perturbation theory, loop diagrams involve the Goldstone bosons. Do they
accurately reflect the effect of $W_L$ loops? We do not have a general answer
here. However, for our process we are able to assure ourselves that such use is
allowed. This is because we are able to reformulate the problem as a dispersion
relation, which only involves external Goldstone bosons. The ingredients to the
dispersion relation are the \(\gamma W_L\rightarrow\gamma W_L\) and the
\(W_L^+W_L^-\rightarrow W_L^+W_L^-\) scattering amplitudes and the substraction
constants. All can be defined for external particles. However in~\cite{2} it
was shown how the choice of the lowest order chiral amplitudes for these
quantities exactly reproduces the result of the set of Feynmann diagrams in
chiral perturbation theory.

The only possible remaining problem could be the sensitivity to the low
momentum region inside the loop integrals, where there is no equivalence
between $W_L$ and the Goldstone bosons. If the loops were infrared dominated,
the infrared contribution would not be trustworthy. Fortunately chiral
amplitudes vanish at zero momentum, so that the infrared region is not too
important. This also can seen by the fact that the factor of \(\ln{m_\pi^2}\)
vanish and the result has a smooth limit as $m_\pi\rightarrow 0$. We have also
checked that truncating the dispersion integrals at low energies has little
effect on the resulting cross sections.

\newpage

\pagenumbering{roman}
\setlength{\unitlength}{0.16pt}
\begin{center}

\end{center}
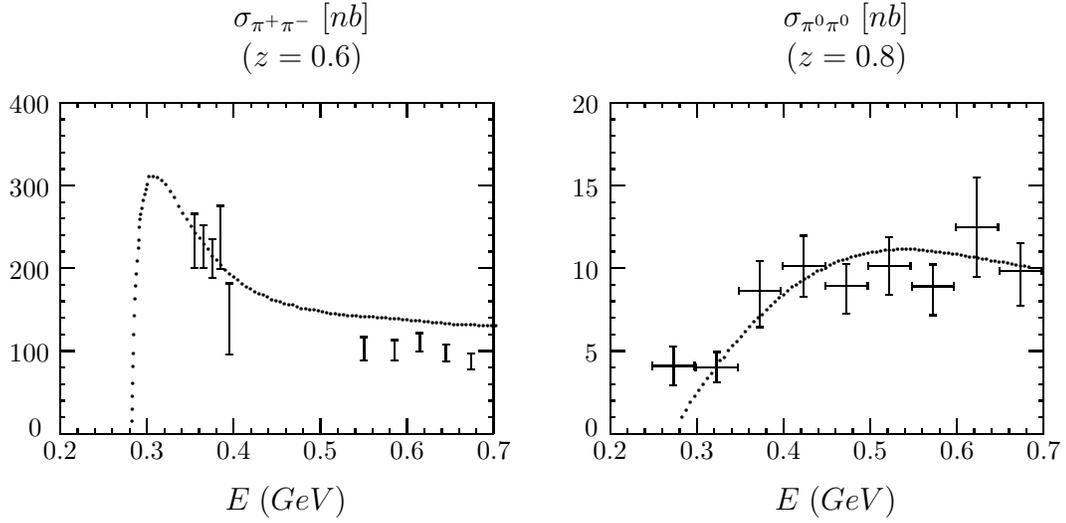
{\bf Figure~1.} Total $\gamma\gamma\rightarrow\pi^+\pi^-$ and
$\gamma\gamma\rightarrow\pi^0\pi^0$ cross sections with a forward\\ cut at
$\mid\!z\!\mid\,\equiv\,\mid\!\cos \Theta\!\mid$, figures taken
from~\cite{2}.

\setlength{\unitlength}{0.240900pt}

\begin{center}
%
\end{center}
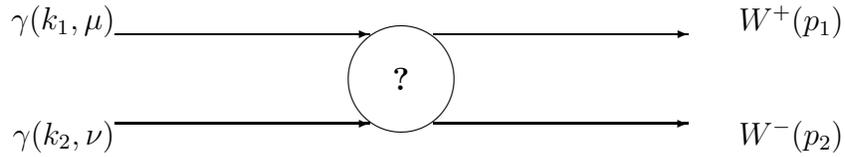
{\bf Figure~2.} Explanation of the kinematic
conventions used in the text.
\newpage

\setlength{\unitlength}{0.24pt}
\ifx\plotpoint\undefined\newsavebox{\plotpoint}\fi
\sbox{\plotpoint}{\rule[-0.175pt]{0.350pt}{0.350pt}}%
\begin{center}
%
\end{center}
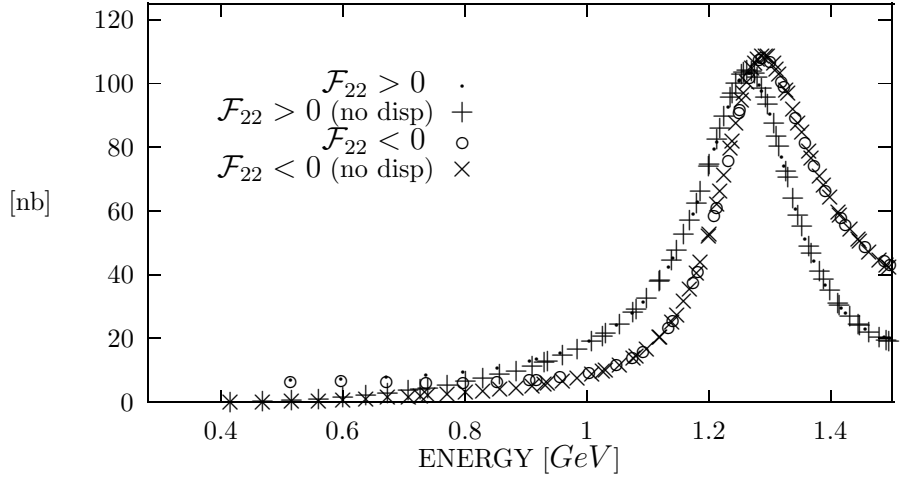
{\bf Figure~3.} $\gamma\gamma\rightarrow\pi^0\pi^0$ cross section with forward
cut at $\mid Z\mid\equiv\mid \cos \Theta\mid<.6$.

\begin{center}
%
\end{center}
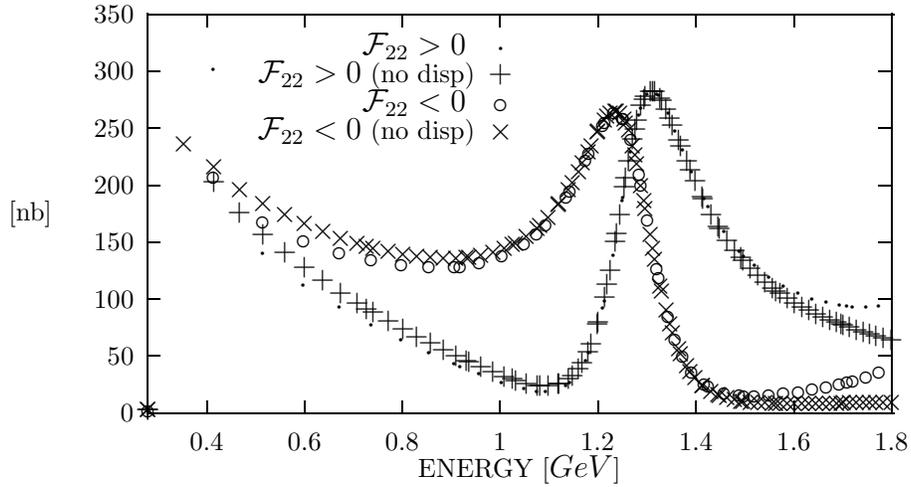
{\bf Figure~4.} $\gamma\gamma\rightarrow\pi^+\pi^-$ cross section with forward
cut at $\mid Z\mid\equiv\mid \cos \Theta\mid<.6$. Note that at $f_2$ resonance
the dispersion integral is unimportant (`no disp' means that the dispersion
integral and the substraction terms are left out).

\newpage
\setlength{\unitlength}{0.240900pt}

\begin{center}
%
\end{center}
{\bf Figure~5.} $1/N$ power counting diagrams for $f_2$ interactions.

\begin{center}
%
\end{center}
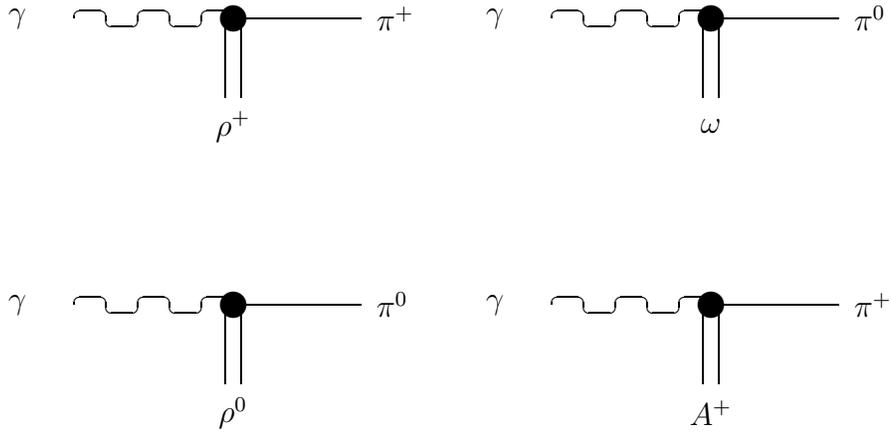
{\bf Figure~6.} Vector meson -- photon -- pion
vertices: these pick up form factors
at $t\geq m^2_\rho$.

\newpage
\setlength{\unitlength}{0.26pt}
\ifx\plotpoint\undefined\newsavebox{\plotpoint}\fi
\sbox{\plotpoint}{\rule[-0.175pt]{0.350pt}{0.350pt}}%
\begin{center}
%

\end{center}
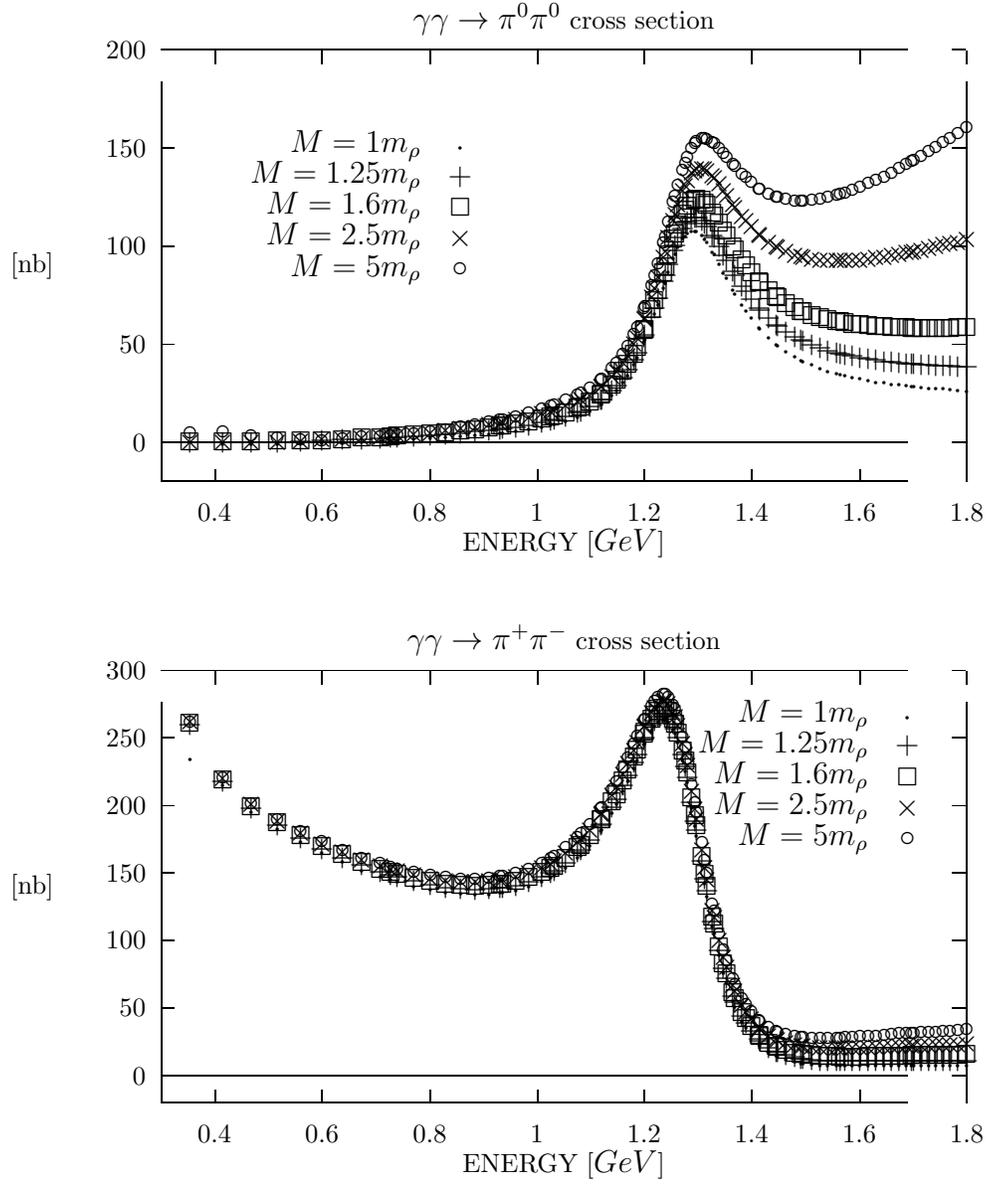
{\bf Figure~7~a,b.} The impact of the form factors on
$\gamma\gamma\rightarrow\pi\pi$ cross sections.

\newpage

\setlength{\unitlength}{0.26pt}
\ifx\plotpoint\undefined\newsavebox{\plotpoint}\fi
\sbox{\plotpoint}{\rule[-0.175pt]{0.350pt}{0.350pt}}%
\begin{center}
%

\end{center}
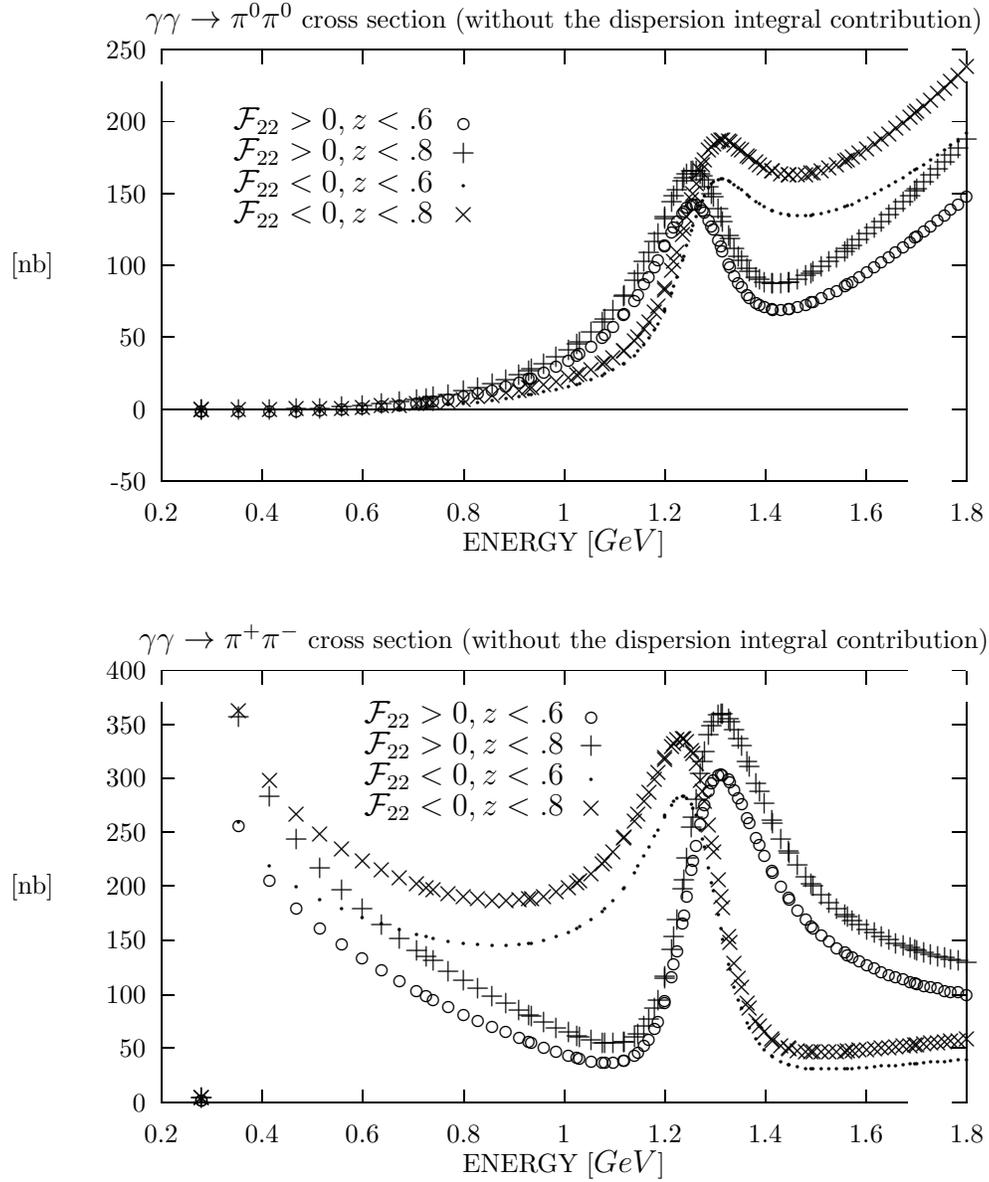
{\bf Figure~8~a,b.} Pointlike vector mesons would generate an
increase in $\gamma\gamma\rightarrow\pi\pi$ cross sections and
would even make it hard to determine the sign of
${\cal F}_{22}$ .
\newpage

\setlength{\unitlength}{0.240900pt}
\ifx\plotpoint\undefined\newsavebox{\plotpoint}\fi
\sbox{\plotpoint}{\rule[-0.175pt]{0.350pt}{0.350pt}}%
\begin{center}
%

\end{center}
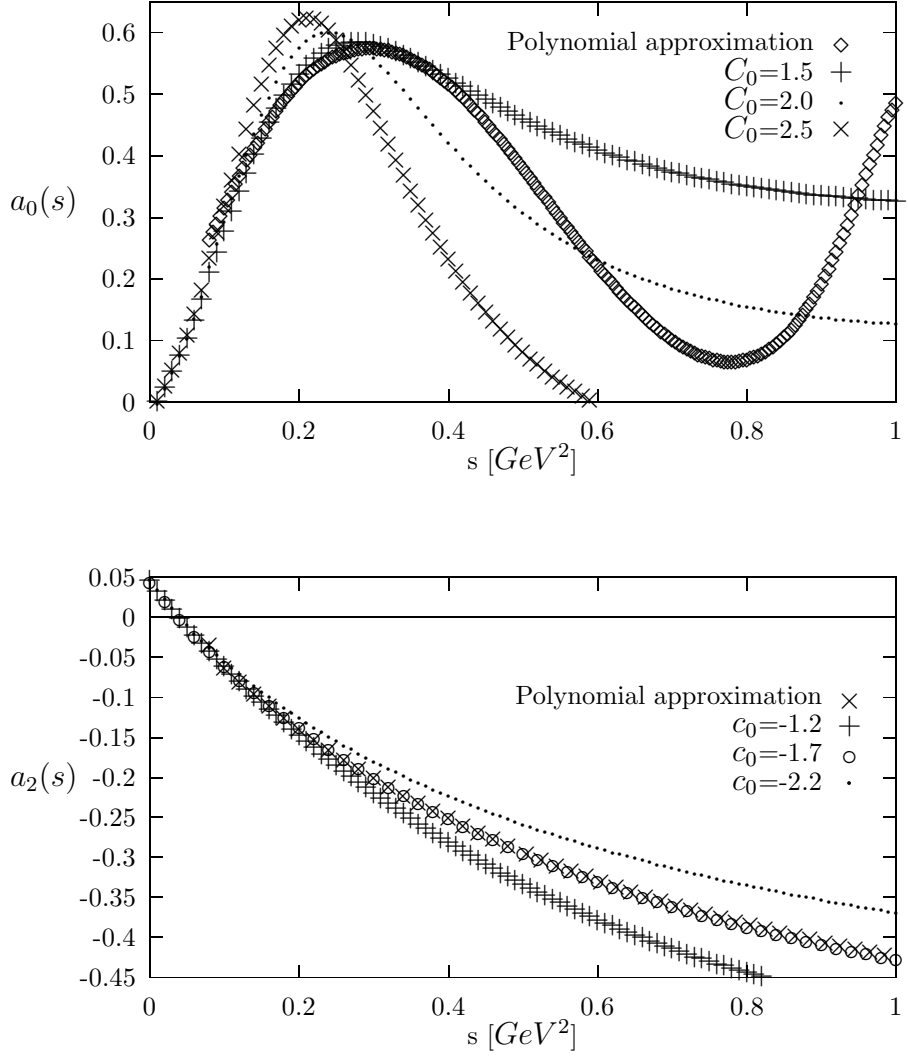
{\bf Figure~9~a,b.} Pad\'e approximation to the real part of the isospin-0 and
isospin-2 $\pi\pi$ elastic scattering amplitude. The curves signed 'polynomial
approximation` are a) for $I=0$ fitted Gasser's data~\cite{gas14}, corrected
to have correct chiral behaviour at low $s$, b) fitted similarly to Pad\'e
approximants of~\cite{2} and to the linear chiral prediction.

\newpage

\setlength{\unitlength}{0.240900pt}
\ifx\plotpoint\undefined\newsavebox{\plotpoint}\fi
\sbox{\plotpoint}{\rule[-0.175pt]{0.350pt}{0.350pt}}%
\begin{center}
\begin{picture}(1500,900)(0,0)
\tenrm
\sbox{\plotpoint}{\rule[-0.175pt]{0.350pt}{0.350pt}}%
\put(264,158){\rule[-0.175pt]{0.350pt}{151.526pt}}
\put(264,158){\rule[-0.175pt]{2.409pt}{0.350pt}}
\put(1426,158){\rule[-0.175pt]{2.409pt}{0.350pt}}
\put(264,172){\rule[-0.175pt]{4.818pt}{0.350pt}}
\put(242,172){\makebox(0,0)[r]{0.01}}
\put(1416,172){\rule[-0.175pt]{4.818pt}{0.350pt}}
\put(264,265){\rule[-0.175pt]{2.409pt}{0.350pt}}
\put(1426,265){\rule[-0.175pt]{2.409pt}{0.350pt}}
\put(264,319){\rule[-0.175pt]{2.409pt}{0.350pt}}
\put(1426,319){\rule[-0.175pt]{2.409pt}{0.350pt}}
\put(264,357){\rule[-0.175pt]{2.409pt}{0.350pt}}
\put(1426,357){\rule[-0.175pt]{2.409pt}{0.350pt}}
\put(264,387){\rule[-0.175pt]{2.409pt}{0.350pt}}
\put(1426,387){\rule[-0.175pt]{2.409pt}{0.350pt}}
\put(264,411){\rule[-0.175pt]{2.409pt}{0.350pt}}
\put(1426,411){\rule[-0.175pt]{2.409pt}{0.350pt}}
\put(264,432){\rule[-0.175pt]{2.409pt}{0.350pt}}
\put(1426,432){\rule[-0.175pt]{2.409pt}{0.350pt}}
\put(264,450){\rule[-0.175pt]{2.409pt}{0.350pt}}
\put(1426,450){\rule[-0.175pt]{2.409pt}{0.350pt}}
\put(264,465){\rule[-0.175pt]{2.409pt}{0.350pt}}
\put(1426,465){\rule[-0.175pt]{2.409pt}{0.350pt}}
\put(264,480){\rule[-0.175pt]{4.818pt}{0.350pt}}
\put(242,480){\makebox(0,0)[r]{0.1}}
\put(1416,480){\rule[-0.175pt]{4.818pt}{0.350pt}}
\put(264,572){\rule[-0.175pt]{2.409pt}{0.350pt}}
\put(1426,572){\rule[-0.175pt]{2.409pt}{0.350pt}}
\put(264,626){\rule[-0.175pt]{2.409pt}{0.350pt}}
\put(1426,626){\rule[-0.175pt]{2.409pt}{0.350pt}}
\put(264,665){\rule[-0.175pt]{2.409pt}{0.350pt}}
\put(1426,665){\rule[-0.175pt]{2.409pt}{0.350pt}}
\put(264,694){\rule[-0.175pt]{2.409pt}{0.350pt}}
\put(1426,694){\rule[-0.175pt]{2.409pt}{0.350pt}}
\put(264,719){\rule[-0.175pt]{2.409pt}{0.350pt}}
\put(1426,719){\rule[-0.175pt]{2.409pt}{0.350pt}}
\put(264,739){\rule[-0.175pt]{2.409pt}{0.350pt}}
\put(1426,739){\rule[-0.175pt]{2.409pt}{0.350pt}}
\put(264,757){\rule[-0.175pt]{2.409pt}{0.350pt}}
\put(1426,757){\rule[-0.175pt]{2.409pt}{0.350pt}}
\put(264,773){\rule[-0.175pt]{2.409pt}{0.350pt}}
\put(1426,773){\rule[-0.175pt]{2.409pt}{0.350pt}}
\put(264,787){\rule[-0.175pt]{4.818pt}{0.350pt}}
\put(242,787){\makebox(0,0)[r]{1}}
\put(1416,787){\rule[-0.175pt]{4.818pt}{0.350pt}}
\put(264,158){\rule[-0.175pt]{0.350pt}{4.818pt}}
\put(264,113){\makebox(0,0){0}}
\put(264,767){\rule[-0.175pt]{0.350pt}{4.818pt}}
\put(411,158){\rule[-0.175pt]{0.350pt}{4.818pt}}
\put(411,113){\makebox(0,0){500}}
\put(411,767){\rule[-0.175pt]{0.350pt}{4.818pt}}
\put(557,158){\rule[-0.175pt]{0.350pt}{4.818pt}}
\put(557,113){\makebox(0,0){1000}}
\put(557,767){\rule[-0.175pt]{0.350pt}{4.818pt}}
\put(704,158){\rule[-0.175pt]{0.350pt}{4.818pt}}
\put(704,113){\makebox(0,0){1500}}
\put(704,767){\rule[-0.175pt]{0.350pt}{4.818pt}}
\put(850,158){\rule[-0.175pt]{0.350pt}{4.818pt}}
\put(850,113){\makebox(0,0){2000}}
\put(850,767){\rule[-0.175pt]{0.350pt}{4.818pt}}
\put(997,158){\rule[-0.175pt]{0.350pt}{4.818pt}}
\put(997,113){\makebox(0,0){2500}}
\put(997,767){\rule[-0.175pt]{0.350pt}{4.818pt}}
\put(1143,158){\rule[-0.175pt]{0.350pt}{4.818pt}}
\put(1143,113){\makebox(0,0){3000}}
\put(1143,767){\rule[-0.175pt]{0.350pt}{4.818pt}}
\put(1290,158){\rule[-0.175pt]{0.350pt}{4.818pt}}
\put(1290,113){\makebox(0,0){3500}}
\put(1290,767){\rule[-0.175pt]{0.350pt}{4.818pt}}
\put(1436,158){\rule[-0.175pt]{0.350pt}{4.818pt}}
\put(1436,113){\makebox(0,0){4000}}
\put(1436,767){\rule[-0.175pt]{0.350pt}{4.818pt}}
\put(264,158){\rule[-0.175pt]{282.335pt}{0.350pt}}
\put(1436,158){\rule[-0.175pt]{0.350pt}{151.526pt}}
\put(264,787){\rule[-0.175pt]{282.335pt}{0.350pt}}
\put(45,472){\makebox(0,0)[l]{\shortstack{[pb]}}}
\put(850,68){\makebox(0,0){ENERGY [$GeV$]}}
\put(850,832){\makebox(0,0){$\gamma\gamma \rightarrow W^+_LW^-_L$cross
section}}
\put(264,158){\rule[-0.175pt]{0.350pt}{151.526pt}}
\put(1306,722){\makebox(0,0)[r]{$N_{TC}$=3}}
\put(1350,722){\raisebox{-1.2pt}{\makebox(0,0){$\Diamond$}}}
\put(323,734){\raisebox{-1.2pt}{\makebox(0,0){$\Diamond$}}}
\put(326,728){\raisebox{-1.2pt}{\makebox(0,0){$\Diamond$}}}
\put(329,719){\raisebox{-1.2pt}{\makebox(0,0){$\Diamond$}}}
\put(332,709){\raisebox{-1.2pt}{\makebox(0,0){$\Diamond$}}}
\put(335,701){\raisebox{-1.2pt}{\makebox(0,0){$\Diamond$}}}
\put(338,693){\raisebox{-1.2pt}{\makebox(0,0){$\Diamond$}}}
\put(341,687){\raisebox{-1.2pt}{\makebox(0,0){$\Diamond$}}}
\put(344,681){\raisebox{-1.2pt}{\makebox(0,0){$\Diamond$}}}
\put(347,676){\raisebox{-1.2pt}{\makebox(0,0){$\Diamond$}}}
\put(350,672){\raisebox{-1.2pt}{\makebox(0,0){$\Diamond$}}}
\put(354,667){\raisebox{-1.2pt}{\makebox(0,0){$\Diamond$}}}
\put(357,663){\raisebox{-1.2pt}{\makebox(0,0){$\Diamond$}}}
\put(360,659){\raisebox{-1.2pt}{\makebox(0,0){$\Diamond$}}}
\put(363,656){\raisebox{-1.2pt}{\makebox(0,0){$\Diamond$}}}
\put(366,652){\raisebox{-1.2pt}{\makebox(0,0){$\Diamond$}}}
\put(369,648){\raisebox{-1.2pt}{\makebox(0,0){$\Diamond$}}}
\put(372,645){\raisebox{-1.2pt}{\makebox(0,0){$\Diamond$}}}
\put(375,641){\raisebox{-1.2pt}{\makebox(0,0){$\Diamond$}}}
\put(378,637){\raisebox{-1.2pt}{\makebox(0,0){$\Diamond$}}}
\put(381,633){\raisebox{-1.2pt}{\makebox(0,0){$\Diamond$}}}
\put(384,630){\raisebox{-1.2pt}{\makebox(0,0){$\Diamond$}}}
\put(387,626){\raisebox{-1.2pt}{\makebox(0,0){$\Diamond$}}}
\put(390,622){\raisebox{-1.2pt}{\makebox(0,0){$\Diamond$}}}
\put(393,618){\raisebox{-1.2pt}{\makebox(0,0){$\Diamond$}}}
\put(396,614){\raisebox{-1.2pt}{\makebox(0,0){$\Diamond$}}}
\put(399,611){\raisebox{-1.2pt}{\makebox(0,0){$\Diamond$}}}
\put(403,607){\raisebox{-1.2pt}{\makebox(0,0){$\Diamond$}}}
\put(406,603){\raisebox{-1.2pt}{\makebox(0,0){$\Diamond$}}}
\put(409,599){\raisebox{-1.2pt}{\makebox(0,0){$\Diamond$}}}
\put(412,596){\raisebox{-1.2pt}{\makebox(0,0){$\Diamond$}}}
\put(415,592){\raisebox{-1.2pt}{\makebox(0,0){$\Diamond$}}}
\put(418,588){\raisebox{-1.2pt}{\makebox(0,0){$\Diamond$}}}
\put(438,564){\raisebox{-1.2pt}{\makebox(0,0){$\Diamond$}}}
\put(466,534){\raisebox{-1.2pt}{\makebox(0,0){$\Diamond$}}}
\put(493,506){\raisebox{-1.2pt}{\makebox(0,0){$\Diamond$}}}
\put(521,481){\raisebox{-1.2pt}{\makebox(0,0){$\Diamond$}}}
\put(549,457){\raisebox{-1.2pt}{\makebox(0,0){$\Diamond$}}}
\put(576,436){\raisebox{-1.2pt}{\makebox(0,0){$\Diamond$}}}
\put(604,417){\raisebox{-1.2pt}{\makebox(0,0){$\Diamond$}}}
\put(631,399){\raisebox{-1.2pt}{\makebox(0,0){$\Diamond$}}}
\put(659,383){\raisebox{-1.2pt}{\makebox(0,0){$\Diamond$}}}
\put(687,368){\raisebox{-1.2pt}{\makebox(0,0){$\Diamond$}}}
\put(714,355){\raisebox{-1.2pt}{\makebox(0,0){$\Diamond$}}}
\put(742,342){\raisebox{-1.2pt}{\makebox(0,0){$\Diamond$}}}
\put(769,331){\raisebox{-1.2pt}{\makebox(0,0){$\Diamond$}}}
\put(797,321){\raisebox{-1.2pt}{\makebox(0,0){$\Diamond$}}}
\put(824,312){\raisebox{-1.2pt}{\makebox(0,0){$\Diamond$}}}
\put(852,304){\raisebox{-1.2pt}{\makebox(0,0){$\Diamond$}}}
\put(879,297){\raisebox{-1.2pt}{\makebox(0,0){$\Diamond$}}}
\put(907,292){\raisebox{-1.2pt}{\makebox(0,0){$\Diamond$}}}
\put(935,288){\raisebox{-1.2pt}{\makebox(0,0){$\Diamond$}}}
\put(962,285){\raisebox{-1.2pt}{\makebox(0,0){$\Diamond$}}}
\put(990,284){\raisebox{-1.2pt}{\makebox(0,0){$\Diamond$}}}
\put(1017,285){\raisebox{-1.2pt}{\makebox(0,0){$\Diamond$}}}
\put(1045,287){\raisebox{-1.2pt}{\makebox(0,0){$\Diamond$}}}
\put(1072,293){\raisebox{-1.2pt}{\makebox(0,0){$\Diamond$}}}
\put(1100,301){\raisebox{-1.2pt}{\makebox(0,0){$\Diamond$}}}
\put(1127,312){\raisebox{-1.2pt}{\makebox(0,0){$\Diamond$}}}
\put(1155,327){\raisebox{-1.2pt}{\makebox(0,0){$\Diamond$}}}
\put(1183,344){\raisebox{-1.2pt}{\makebox(0,0){$\Diamond$}}}
\put(1210,360){\raisebox{-1.2pt}{\makebox(0,0){$\Diamond$}}}
\put(1238,363){\raisebox{-1.2pt}{\makebox(0,0){$\Diamond$}}}
\put(1265,336){\raisebox{-1.2pt}{\makebox(0,0){$\Diamond$}}}
\put(1293,273){\raisebox{-1.2pt}{\makebox(0,0){$\Diamond$}}}
\put(1321,196){\raisebox{-1.2pt}{\makebox(0,0){$\Diamond$}}}
\put(832,310){\raisebox{-1.2pt}{\makebox(0,0){$\Diamond$}}}
\put(991,284){\raisebox{-1.2pt}{\makebox(0,0){$\Diamond$}}}
\put(1065,291){\raisebox{-1.2pt}{\makebox(0,0){$\Diamond$}}}
\put(1109,304){\raisebox{-1.2pt}{\makebox(0,0){$\Diamond$}}}
\put(1138,318){\raisebox{-1.2pt}{\makebox(0,0){$\Diamond$}}}
\put(1160,330){\raisebox{-1.2pt}{\makebox(0,0){$\Diamond$}}}
\put(1177,341){\raisebox{-1.2pt}{\makebox(0,0){$\Diamond$}}}
\put(1191,350){\raisebox{-1.2pt}{\makebox(0,0){$\Diamond$}}}
\put(1202,356){\raisebox{-1.2pt}{\makebox(0,0){$\Diamond$}}}
\put(1212,361){\raisebox{-1.2pt}{\makebox(0,0){$\Diamond$}}}
\put(1221,364){\raisebox{-1.2pt}{\makebox(0,0){$\Diamond$}}}
\put(1229,365){\raisebox{-1.2pt}{\makebox(0,0){$\Diamond$}}}
\put(1236,364){\raisebox{-1.2pt}{\makebox(0,0){$\Diamond$}}}
\put(1243,361){\raisebox{-1.2pt}{\makebox(0,0){$\Diamond$}}}
\put(1250,356){\raisebox{-1.2pt}{\makebox(0,0){$\Diamond$}}}
\put(1257,348){\raisebox{-1.2pt}{\makebox(0,0){$\Diamond$}}}
\put(1263,339){\raisebox{-1.2pt}{\makebox(0,0){$\Diamond$}}}
\put(1270,327){\raisebox{-1.2pt}{\makebox(0,0){$\Diamond$}}}
\put(1277,313){\raisebox{-1.2pt}{\makebox(0,0){$\Diamond$}}}
\put(1284,296){\raisebox{-1.2pt}{\makebox(0,0){$\Diamond$}}}
\put(1292,275){\raisebox{-1.2pt}{\makebox(0,0){$\Diamond$}}}
\put(1300,252){\raisebox{-1.2pt}{\makebox(0,0){$\Diamond$}}}
\put(1310,226){\raisebox{-1.2pt}{\makebox(0,0){$\Diamond$}}}
\put(1321,196){\raisebox{-1.2pt}{\makebox(0,0){$\Diamond$}}}
\put(1333,163){\raisebox{-1.2pt}{\makebox(0,0){$\Diamond$}}}
\sbox{\plotpoint}{\rule[-0.350pt]{0.700pt}{0.700pt}}%
\put(1306,677){\makebox(0,0)[r]{$N_{TC}$=6}}
\put(1350,677){\makebox(0,0){$+$}}
\put(321,744){\makebox(0,0){$+$}}
\put(323,746){\makebox(0,0){$+$}}
\put(326,742){\makebox(0,0){$+$}}
\put(328,736){\makebox(0,0){$+$}}
\put(330,730){\makebox(0,0){$+$}}
\put(332,724){\makebox(0,0){$+$}}
\put(335,718){\makebox(0,0){$+$}}
\put(337,712){\makebox(0,0){$+$}}
\put(339,708){\makebox(0,0){$+$}}
\put(341,703){\makebox(0,0){$+$}}
\put(344,699){\makebox(0,0){$+$}}
\put(346,695){\makebox(0,0){$+$}}
\put(348,691){\makebox(0,0){$+$}}
\put(350,688){\makebox(0,0){$+$}}
\put(353,685){\makebox(0,0){$+$}}
\put(355,682){\makebox(0,0){$+$}}
\put(357,678){\makebox(0,0){$+$}}
\put(359,675){\makebox(0,0){$+$}}
\put(361,673){\makebox(0,0){$+$}}
\put(364,670){\makebox(0,0){$+$}}
\put(366,667){\makebox(0,0){$+$}}
\put(368,664){\makebox(0,0){$+$}}
\put(370,661){\makebox(0,0){$+$}}
\put(373,658){\makebox(0,0){$+$}}
\put(375,655){\makebox(0,0){$+$}}
\put(377,652){\makebox(0,0){$+$}}
\put(379,650){\makebox(0,0){$+$}}
\put(382,647){\makebox(0,0){$+$}}
\put(384,644){\makebox(0,0){$+$}}
\put(386,641){\makebox(0,0){$+$}}
\put(388,638){\makebox(0,0){$+$}}
\put(391,635){\makebox(0,0){$+$}}
\put(405,617){\makebox(0,0){$+$}}
\put(426,592){\makebox(0,0){$+$}}
\put(446,569){\makebox(0,0){$+$}}
\put(466,548){\makebox(0,0){$+$}}
\put(486,528){\makebox(0,0){$+$}}
\put(506,510){\makebox(0,0){$+$}}
\put(526,493){\makebox(0,0){$+$}}
\put(546,478){\makebox(0,0){$+$}}
\put(566,464){\makebox(0,0){$+$}}
\put(587,451){\makebox(0,0){$+$}}
\put(607,440){\makebox(0,0){$+$}}
\put(627,430){\makebox(0,0){$+$}}
\put(647,421){\makebox(0,0){$+$}}
\put(667,414){\makebox(0,0){$+$}}
\put(687,408){\makebox(0,0){$+$}}
\put(707,403){\makebox(0,0){$+$}}
\put(728,400){\makebox(0,0){$+$}}
\put(748,398){\makebox(0,0){$+$}}
\put(768,398){\makebox(0,0){$+$}}
\put(788,400){\makebox(0,0){$+$}}
\put(808,404){\makebox(0,0){$+$}}
\put(828,411){\makebox(0,0){$+$}}
\put(848,422){\makebox(0,0){$+$}}
\put(868,437){\makebox(0,0){$+$}}
\put(889,459){\makebox(0,0){$+$}}
\put(909,491){\makebox(0,0){$+$}}
\put(929,538){\makebox(0,0){$+$}}
\put(949,596){\makebox(0,0){$+$}}
\put(969,597){\makebox(0,0){$+$}}
\put(989,480){\makebox(0,0){$+$}}
\put(1009,361){\makebox(0,0){$+$}}
\put(1030,290){\makebox(0,0){$+$}}
\put(1050,260){\makebox(0,0){$+$}}
\put(711,403){\makebox(0,0){$+$}}
\put(835,415){\makebox(0,0){$+$}}
\put(879,447){\makebox(0,0){$+$}}
\put(901,477){\makebox(0,0){$+$}}
\put(915,504){\makebox(0,0){$+$}}
\put(925,528){\makebox(0,0){$+$}}
\put(932,548){\makebox(0,0){$+$}}
\put(938,564){\makebox(0,0){$+$}}
\put(943,578){\makebox(0,0){$+$}}
\put(947,590){\makebox(0,0){$+$}}
\put(950,599){\makebox(0,0){$+$}}
\put(953,606){\makebox(0,0){$+$}}
\put(956,610){\makebox(0,0){$+$}}
\put(958,613){\makebox(0,0){$+$}}
\put(961,613){\makebox(0,0){$+$}}
\put(963,611){\makebox(0,0){$+$}}
\put(966,607){\makebox(0,0){$+$}}
\put(968,601){\makebox(0,0){$+$}}
\put(970,593){\makebox(0,0){$+$}}
\put(973,582){\makebox(0,0){$+$}}
\put(975,569){\makebox(0,0){$+$}}
\put(978,553){\makebox(0,0){$+$}}
\put(981,533){\makebox(0,0){$+$}}
\put(985,511){\makebox(0,0){$+$}}
\put(989,485){\makebox(0,0){$+$}}
\put(993,454){\makebox(0,0){$+$}}
\put(999,419){\makebox(0,0){$+$}}
\put(1006,379){\makebox(0,0){$+$}}
\put(1015,335){\makebox(0,0){$+$}}
\put(1029,291){\makebox(0,0){$+$}}
\put(1051,259){\makebox(0,0){$+$}}
\put(1070,252){\makebox(0,0){$+$}}
\put(1070,252){\makebox(0,0){$+$}}
\sbox{\plotpoint}{\rule[-0.500pt]{1.000pt}{1.000pt}}%
\put(1306,632){\makebox(0,0)[r]{$N_{TC}$=9}}
\put(1350,632){\raisebox{-1.2pt}{\makebox(0,0){$\Box$}}}
\put(321,753){\raisebox{-1.2pt}{\makebox(0,0){$\Box$}}}
\put(323,757){\raisebox{-1.2pt}{\makebox(0,0){$\Box$}}}
\put(325,756){\raisebox{-1.2pt}{\makebox(0,0){$\Box$}}}
\put(327,752){\raisebox{-1.2pt}{\makebox(0,0){$\Box$}}}
\put(329,747){\raisebox{-1.2pt}{\makebox(0,0){$\Box$}}}
\put(331,742){\raisebox{-1.2pt}{\makebox(0,0){$\Box$}}}
\put(333,737){\raisebox{-1.2pt}{\makebox(0,0){$\Box$}}}
\put(335,732){\raisebox{-1.2pt}{\makebox(0,0){$\Box$}}}
\put(337,727){\raisebox{-1.2pt}{\makebox(0,0){$\Box$}}}
\put(339,723){\raisebox{-1.2pt}{\makebox(0,0){$\Box$}}}
\put(341,719){\raisebox{-1.2pt}{\makebox(0,0){$\Box$}}}
\put(343,715){\raisebox{-1.2pt}{\makebox(0,0){$\Box$}}}
\put(345,712){\raisebox{-1.2pt}{\makebox(0,0){$\Box$}}}
\put(347,708){\raisebox{-1.2pt}{\makebox(0,0){$\Box$}}}
\put(349,705){\raisebox{-1.2pt}{\makebox(0,0){$\Box$}}}
\put(351,702){\raisebox{-1.2pt}{\makebox(0,0){$\Box$}}}
\put(353,699){\raisebox{-1.2pt}{\makebox(0,0){$\Box$}}}
\put(355,696){\raisebox{-1.2pt}{\makebox(0,0){$\Box$}}}
\put(357,693){\raisebox{-1.2pt}{\makebox(0,0){$\Box$}}}
\put(359,690){\raisebox{-1.2pt}{\makebox(0,0){$\Box$}}}
\put(361,687){\raisebox{-1.2pt}{\makebox(0,0){$\Box$}}}
\put(363,685){\raisebox{-1.2pt}{\makebox(0,0){$\Box$}}}
\put(365,682){\raisebox{-1.2pt}{\makebox(0,0){$\Box$}}}
\put(367,679){\raisebox{-1.2pt}{\makebox(0,0){$\Box$}}}
\put(369,677){\raisebox{-1.2pt}{\makebox(0,0){$\Box$}}}
\put(371,674){\raisebox{-1.2pt}{\makebox(0,0){$\Box$}}}
\put(373,671){\raisebox{-1.2pt}{\makebox(0,0){$\Box$}}}
\put(375,669){\raisebox{-1.2pt}{\makebox(0,0){$\Box$}}}
\put(377,665){\raisebox{-1.2pt}{\makebox(0,0){$\Box$}}}
\put(395,642){\raisebox{-1.2pt}{\makebox(0,0){$\Box$}}}
\put(413,620){\raisebox{-1.2pt}{\makebox(0,0){$\Box$}}}
\put(431,598){\raisebox{-1.2pt}{\makebox(0,0){$\Box$}}}
\put(449,579){\raisebox{-1.2pt}{\makebox(0,0){$\Box$}}}
\put(467,560){\raisebox{-1.2pt}{\makebox(0,0){$\Box$}}}
\put(485,544){\raisebox{-1.2pt}{\makebox(0,0){$\Box$}}}
\put(503,529){\raisebox{-1.2pt}{\makebox(0,0){$\Box$}}}
\put(521,515){\raisebox{-1.2pt}{\makebox(0,0){$\Box$}}}
\put(539,504){\raisebox{-1.2pt}{\makebox(0,0){$\Box$}}}
\put(557,493){\raisebox{-1.2pt}{\makebox(0,0){$\Box$}}}
\put(575,485){\raisebox{-1.2pt}{\makebox(0,0){$\Box$}}}
\put(593,478){\raisebox{-1.2pt}{\makebox(0,0){$\Box$}}}
\put(611,473){\raisebox{-1.2pt}{\makebox(0,0){$\Box$}}}
\put(629,470){\raisebox{-1.2pt}{\makebox(0,0){$\Box$}}}
\put(647,469){\raisebox{-1.2pt}{\makebox(0,0){$\Box$}}}
\put(665,469){\raisebox{-1.2pt}{\makebox(0,0){$\Box$}}}
\put(683,471){\raisebox{-1.2pt}{\makebox(0,0){$\Box$}}}
\put(701,476){\raisebox{-1.2pt}{\makebox(0,0){$\Box$}}}
\put(719,484){\raisebox{-1.2pt}{\makebox(0,0){$\Box$}}}
\put(737,495){\raisebox{-1.2pt}{\makebox(0,0){$\Box$}}}
\put(755,511){\raisebox{-1.2pt}{\makebox(0,0){$\Box$}}}
\put(773,534){\raisebox{-1.2pt}{\makebox(0,0){$\Box$}}}
\put(791,572){\raisebox{-1.2pt}{\makebox(0,0){$\Box$}}}
\put(809,636){\raisebox{-1.2pt}{\makebox(0,0){$\Box$}}}
\put(827,742){\raisebox{-1.2pt}{\makebox(0,0){$\Box$}}}
\put(845,701){\raisebox{-1.2pt}{\makebox(0,0){$\Box$}}}
\put(863,521){\raisebox{-1.2pt}{\makebox(0,0){$\Box$}}}
\put(881,434){\raisebox{-1.2pt}{\makebox(0,0){$\Box$}}}
\put(899,408){\raisebox{-1.2pt}{\makebox(0,0){$\Box$}}}
\put(917,405){\raisebox{-1.2pt}{\makebox(0,0){$\Box$}}}
\put(678,470){\raisebox{-1.2pt}{\makebox(0,0){$\Box$}}}
\put(764,521){\raisebox{-1.2pt}{\makebox(0,0){$\Box$}}}
\put(791,570){\raisebox{-1.2pt}{\makebox(0,0){$\Box$}}}
\put(803,610){\raisebox{-1.2pt}{\makebox(0,0){$\Box$}}}
\put(811,644){\raisebox{-1.2pt}{\makebox(0,0){$\Box$}}}
\put(816,673){\raisebox{-1.2pt}{\makebox(0,0){$\Box$}}}
\put(820,697){\raisebox{-1.2pt}{\makebox(0,0){$\Box$}}}
\put(823,716){\raisebox{-1.2pt}{\makebox(0,0){$\Box$}}}
\put(826,732){\raisebox{-1.2pt}{\makebox(0,0){$\Box$}}}
\put(828,745){\raisebox{-1.2pt}{\makebox(0,0){$\Box$}}}
\put(830,755){\raisebox{-1.2pt}{\makebox(0,0){$\Box$}}}
\put(832,762){\raisebox{-1.2pt}{\makebox(0,0){$\Box$}}}
\put(833,766){\raisebox{-1.2pt}{\makebox(0,0){$\Box$}}}
\put(834,768){\raisebox{-1.2pt}{\makebox(0,0){$\Box$}}}
\put(836,767){\raisebox{-1.2pt}{\makebox(0,0){$\Box$}}}
\put(837,763){\raisebox{-1.2pt}{\makebox(0,0){$\Box$}}}
\put(839,757){\raisebox{-1.2pt}{\makebox(0,0){$\Box$}}}
\put(840,748){\raisebox{-1.2pt}{\makebox(0,0){$\Box$}}}
\put(842,736){\raisebox{-1.2pt}{\makebox(0,0){$\Box$}}}
\put(843,720){\raisebox{-1.2pt}{\makebox(0,0){$\Box$}}}
\put(845,702){\raisebox{-1.2pt}{\makebox(0,0){$\Box$}}}
\put(847,679){\raisebox{-1.2pt}{\makebox(0,0){$\Box$}}}
\put(850,652){\raisebox{-1.2pt}{\makebox(0,0){$\Box$}}}
\put(853,620){\raisebox{-1.2pt}{\makebox(0,0){$\Box$}}}
\put(856,582){\raisebox{-1.2pt}{\makebox(0,0){$\Box$}}}
\put(861,538){\raisebox{-1.2pt}{\makebox(0,0){$\Box$}}}
\put(868,489){\raisebox{-1.2pt}{\makebox(0,0){$\Box$}}}
\put(879,439){\raisebox{-1.2pt}{\makebox(0,0){$\Box$}}}
\put(901,407){\raisebox{-1.2pt}{\makebox(0,0){$\Box$}}}
\put(923,406){\raisebox{-1.2pt}{\makebox(0,0){$\Box$}}}
\end{picture}
\end{center}
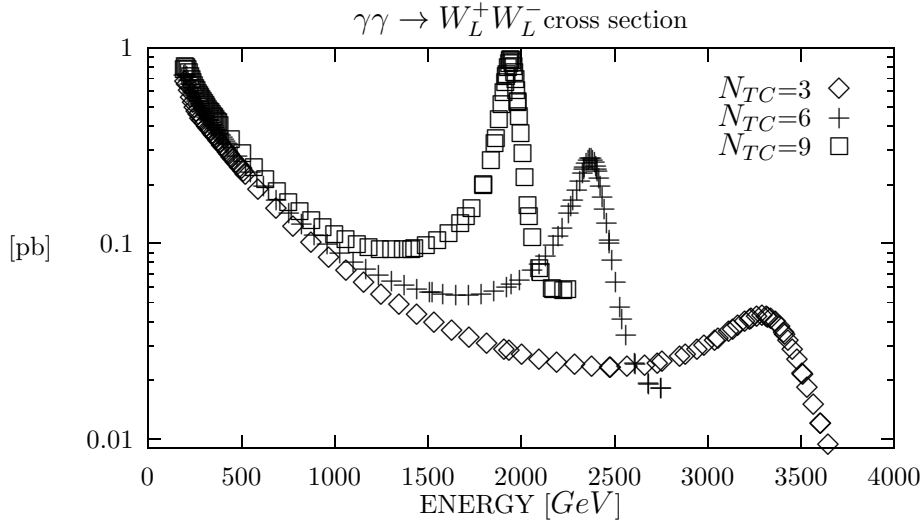
{\bf Figure~10~a.} The total $\gamma\gamma \rightarrow W^+_LW^-_L$ cross
section as a function of $\sqrt{s}$ and $N_{TC}$. Note
the strong dependence of the $f_2$ peak on technicolor number.

\begin{center}
\begin{picture}(1500,900)(0,0)
\tenrm
\sbox{\plotpoint}{\rule[-0.175pt]{0.350pt}{0.350pt}}%
\put(264,158){\rule[-0.175pt]{0.350pt}{151.526pt}}
\put(264,158){\rule[-0.175pt]{4.818pt}{0.350pt}}
\put(242,158){\makebox(0,0)[r]{1}}
\put(1416,158){\rule[-0.175pt]{4.818pt}{0.350pt}}
\put(264,221){\rule[-0.175pt]{2.409pt}{0.350pt}}
\put(1426,221){\rule[-0.175pt]{2.409pt}{0.350pt}}
\put(264,258){\rule[-0.175pt]{2.409pt}{0.350pt}}
\put(1426,258){\rule[-0.175pt]{2.409pt}{0.350pt}}
\put(264,284){\rule[-0.175pt]{2.409pt}{0.350pt}}
\put(1426,284){\rule[-0.175pt]{2.409pt}{0.350pt}}
\put(264,305){\rule[-0.175pt]{2.409pt}{0.350pt}}
\put(1426,305){\rule[-0.175pt]{2.409pt}{0.350pt}}
\put(264,321){\rule[-0.175pt]{2.409pt}{0.350pt}}
\put(1426,321){\rule[-0.175pt]{2.409pt}{0.350pt}}
\put(264,335){\rule[-0.175pt]{2.409pt}{0.350pt}}
\put(1426,335){\rule[-0.175pt]{2.409pt}{0.350pt}}
\put(264,347){\rule[-0.175pt]{2.409pt}{0.350pt}}
\put(1426,347){\rule[-0.175pt]{2.409pt}{0.350pt}}
\put(264,358){\rule[-0.175pt]{2.409pt}{0.350pt}}
\put(1426,358){\rule[-0.175pt]{2.409pt}{0.350pt}}
\put(264,368){\rule[-0.175pt]{4.818pt}{0.350pt}}
\put(242,368){\makebox(0,0)[r]{10}}
\put(1416,368){\rule[-0.175pt]{4.818pt}{0.350pt}}
\put(264,431){\rule[-0.175pt]{2.409pt}{0.350pt}}
\put(1426,431){\rule[-0.175pt]{2.409pt}{0.350pt}}
\put(264,468){\rule[-0.175pt]{2.409pt}{0.350pt}}
\put(1426,468){\rule[-0.175pt]{2.409pt}{0.350pt}}
\put(264,494){\rule[-0.175pt]{2.409pt}{0.350pt}}
\put(1426,494){\rule[-0.175pt]{2.409pt}{0.350pt}}
\put(264,514){\rule[-0.175pt]{2.409pt}{0.350pt}}
\put(1426,514){\rule[-0.175pt]{2.409pt}{0.350pt}}
\put(264,531){\rule[-0.175pt]{2.409pt}{0.350pt}}
\put(1426,531){\rule[-0.175pt]{2.409pt}{0.350pt}}
\put(264,545){\rule[-0.175pt]{2.409pt}{0.350pt}}
\put(1426,545){\rule[-0.175pt]{2.409pt}{0.350pt}}
\put(264,557){\rule[-0.175pt]{2.409pt}{0.350pt}}
\put(1426,557){\rule[-0.175pt]{2.409pt}{0.350pt}}
\put(264,568){\rule[-0.175pt]{2.409pt}{0.350pt}}
\put(1426,568){\rule[-0.175pt]{2.409pt}{0.350pt}}
\put(264,577){\rule[-0.175pt]{4.818pt}{0.350pt}}
\put(242,577){\makebox(0,0)[r]{100}}
\put(1416,577){\rule[-0.175pt]{4.818pt}{0.350pt}}
\put(264,640){\rule[-0.175pt]{2.409pt}{0.350pt}}
\put(1426,640){\rule[-0.175pt]{2.409pt}{0.350pt}}
\put(264,677){\rule[-0.175pt]{2.409pt}{0.350pt}}
\put(1426,677){\rule[-0.175pt]{2.409pt}{0.350pt}}
\put(264,704){\rule[-0.175pt]{2.409pt}{0.350pt}}
\put(1426,704){\rule[-0.175pt]{2.409pt}{0.350pt}}
\put(264,724){\rule[-0.175pt]{2.409pt}{0.350pt}}
\put(1426,724){\rule[-0.175pt]{2.409pt}{0.350pt}}
\put(264,740){\rule[-0.175pt]{2.409pt}{0.350pt}}
\put(1426,740){\rule[-0.175pt]{2.409pt}{0.350pt}}
\put(264,755){\rule[-0.175pt]{2.409pt}{0.350pt}}
\put(1426,755){\rule[-0.175pt]{2.409pt}{0.350pt}}
\put(264,767){\rule[-0.175pt]{2.409pt}{0.350pt}}
\put(1426,767){\rule[-0.175pt]{2.409pt}{0.350pt}}
\put(264,777){\rule[-0.175pt]{2.409pt}{0.350pt}}
\put(1426,777){\rule[-0.175pt]{2.409pt}{0.350pt}}
\put(264,787){\rule[-0.175pt]{4.818pt}{0.350pt}}
\put(242,787){\makebox(0,0)[r]{1000}}
\put(1416,787){\rule[-0.175pt]{4.818pt}{0.350pt}}
\put(264,158){\rule[-0.175pt]{0.350pt}{4.818pt}}
\put(264,113){\makebox(0,0){0}}
\put(264,767){\rule[-0.175pt]{0.350pt}{4.818pt}}
\put(411,158){\rule[-0.175pt]{0.350pt}{4.818pt}}
\put(411,113){\makebox(0,0){500}}
\put(411,767){\rule[-0.175pt]{0.350pt}{4.818pt}}
\put(557,158){\rule[-0.175pt]{0.350pt}{4.818pt}}
\put(557,113){\makebox(0,0){1000}}
\put(557,767){\rule[-0.175pt]{0.350pt}{4.818pt}}
\put(704,158){\rule[-0.175pt]{0.350pt}{4.818pt}}
\put(704,113){\makebox(0,0){1500}}
\put(704,767){\rule[-0.175pt]{0.350pt}{4.818pt}}
\put(850,158){\rule[-0.175pt]{0.350pt}{4.818pt}}
\put(850,113){\makebox(0,0){2000}}
\put(850,767){\rule[-0.175pt]{0.350pt}{4.818pt}}
\put(997,158){\rule[-0.175pt]{0.350pt}{4.818pt}}
\put(997,113){\makebox(0,0){2500}}
\put(997,767){\rule[-0.175pt]{0.350pt}{4.818pt}}
\put(1143,158){\rule[-0.175pt]{0.350pt}{4.818pt}}
\put(1143,113){\makebox(0,0){3000}}
\put(1143,767){\rule[-0.175pt]{0.350pt}{4.818pt}}
\put(1290,158){\rule[-0.175pt]{0.350pt}{4.818pt}}
\put(1290,113){\makebox(0,0){3500}}
\put(1290,767){\rule[-0.175pt]{0.350pt}{4.818pt}}
\put(1436,158){\rule[-0.175pt]{0.350pt}{4.818pt}}
\put(1436,113){\makebox(0,0){4000}}
\put(1436,767){\rule[-0.175pt]{0.350pt}{4.818pt}}
\put(264,158){\rule[-0.175pt]{282.335pt}{0.350pt}}
\put(1436,158){\rule[-0.175pt]{0.350pt}{151.526pt}}
\put(264,787){\rule[-0.175pt]{282.335pt}{0.350pt}}
\put(45,472){\makebox(0,0)[l]{\shortstack{[fb]}}}
\put(850,68){\makebox(0,0){ENERGY [$GeV$]}}
\put(850,832){\makebox(0,0){$\gamma\gamma \rightarrow Z_LZ_L$ cross section}}
\put(264,158){\rule[-0.175pt]{0.350pt}{151.526pt}}
\put(1306,722){\makebox(0,0)[r]{$N_{TC}$=3}}
\put(1350,722){\raisebox{-1.2pt}{\makebox(0,0){$\Diamond$}}}
\put(1072,176){\raisebox{-1.2pt}{\makebox(0,0){$\Diamond$}}}
\put(1100,204){\raisebox{-1.2pt}{\makebox(0,0){$\Diamond$}}}
\put(1127,236){\raisebox{-1.2pt}{\makebox(0,0){$\Diamond$}}}
\put(1155,274){\raisebox{-1.2pt}{\makebox(0,0){$\Diamond$}}}
\put(1183,316){\raisebox{-1.2pt}{\makebox(0,0){$\Diamond$}}}
\put(1210,359){\raisebox{-1.2pt}{\makebox(0,0){$\Diamond$}}}
\put(1238,396){\raisebox{-1.2pt}{\makebox(0,0){$\Diamond$}}}
\put(1265,414){\raisebox{-1.2pt}{\makebox(0,0){$\Diamond$}}}
\put(1293,409){\raisebox{-1.2pt}{\makebox(0,0){$\Diamond$}}}
\put(1321,392){\raisebox{-1.2pt}{\makebox(0,0){$\Diamond$}}}
\put(1065,169){\raisebox{-1.2pt}{\makebox(0,0){$\Diamond$}}}
\put(1109,214){\raisebox{-1.2pt}{\makebox(0,0){$\Diamond$}}}
\put(1138,251){\raisebox{-1.2pt}{\makebox(0,0){$\Diamond$}}}
\put(1160,281){\raisebox{-1.2pt}{\makebox(0,0){$\Diamond$}}}
\put(1177,307){\raisebox{-1.2pt}{\makebox(0,0){$\Diamond$}}}
\put(1191,328){\raisebox{-1.2pt}{\makebox(0,0){$\Diamond$}}}
\put(1202,346){\raisebox{-1.2pt}{\makebox(0,0){$\Diamond$}}}
\put(1212,362){\raisebox{-1.2pt}{\makebox(0,0){$\Diamond$}}}
\put(1221,375){\raisebox{-1.2pt}{\makebox(0,0){$\Diamond$}}}
\put(1229,385){\raisebox{-1.2pt}{\makebox(0,0){$\Diamond$}}}
\put(1236,394){\raisebox{-1.2pt}{\makebox(0,0){$\Diamond$}}}
\put(1243,401){\raisebox{-1.2pt}{\makebox(0,0){$\Diamond$}}}
\put(1250,407){\raisebox{-1.2pt}{\makebox(0,0){$\Diamond$}}}
\put(1257,411){\raisebox{-1.2pt}{\makebox(0,0){$\Diamond$}}}
\put(1263,413){\raisebox{-1.2pt}{\makebox(0,0){$\Diamond$}}}
\put(1270,414){\raisebox{-1.2pt}{\makebox(0,0){$\Diamond$}}}
\put(1277,414){\raisebox{-1.2pt}{\makebox(0,0){$\Diamond$}}}
\put(1284,412){\raisebox{-1.2pt}{\makebox(0,0){$\Diamond$}}}
\put(1292,409){\raisebox{-1.2pt}{\makebox(0,0){$\Diamond$}}}
\put(1300,405){\raisebox{-1.2pt}{\makebox(0,0){$\Diamond$}}}
\put(1310,399){\raisebox{-1.2pt}{\makebox(0,0){$\Diamond$}}}
\put(1321,392){\raisebox{-1.2pt}{\makebox(0,0){$\Diamond$}}}
\put(1333,384){\raisebox{-1.2pt}{\makebox(0,0){$\Diamond$}}}
\put(1348,375){\raisebox{-1.2pt}{\makebox(0,0){$\Diamond$}}}
\put(1348,375){\raisebox{-1.2pt}{\makebox(0,0){$\Diamond$}}}
\put(1348,375){\raisebox{-1.2pt}{\makebox(0,0){$\Diamond$}}}
\put(1348,375){\raisebox{-1.2pt}{\makebox(0,0){$\Diamond$}}}
\put(1348,375){\raisebox{-1.2pt}{\makebox(0,0){$\Diamond$}}}
\put(1348,375){\raisebox{-1.2pt}{\makebox(0,0){$\Diamond$}}}
\put(1348,375){\raisebox{-1.2pt}{\makebox(0,0){$\Diamond$}}}
\sbox{\plotpoint}{\rule[-0.350pt]{0.700pt}{0.700pt}}%
\put(1306,677){\makebox(0,0)[r]{$N_{TC}$=6}}
\put(1350,677){\makebox(0,0){$+$}}
\put(647,160){\makebox(0,0){$+$}}
\put(667,180){\makebox(0,0){$+$}}
\put(687,197){\makebox(0,0){$+$}}
\put(707,211){\makebox(0,0){$+$}}
\put(728,223){\makebox(0,0){$+$}}
\put(748,233){\makebox(0,0){$+$}}
\put(768,242){\makebox(0,0){$+$}}
\put(788,250){\makebox(0,0){$+$}}
\put(808,258){\makebox(0,0){$+$}}
\put(828,267){\makebox(0,0){$+$}}
\put(848,280){\makebox(0,0){$+$}}
\put(868,301){\makebox(0,0){$+$}}
\put(889,335){\makebox(0,0){$+$}}
\put(909,386){\makebox(0,0){$+$}}
\put(929,458){\makebox(0,0){$+$}}
\put(949,546){\makebox(0,0){$+$}}
\put(969,598){\makebox(0,0){$+$}}
\put(989,565){\makebox(0,0){$+$}}
\put(1009,517){\makebox(0,0){$+$}}
\put(1030,481){\makebox(0,0){$+$}}
\put(1050,454){\makebox(0,0){$+$}}
\put(711,213){\makebox(0,0){$+$}}
\put(835,271){\makebox(0,0){$+$}}
\put(879,317){\makebox(0,0){$+$}}
\put(901,365){\makebox(0,0){$+$}}
\put(915,407){\makebox(0,0){$+$}}
\put(925,443){\makebox(0,0){$+$}}
\put(932,473){\makebox(0,0){$+$}}
\put(938,497){\makebox(0,0){$+$}}
\put(943,518){\makebox(0,0){$+$}}
\put(947,536){\makebox(0,0){$+$}}
\put(950,550){\makebox(0,0){$+$}}
\put(953,563){\makebox(0,0){$+$}}
\put(956,573){\makebox(0,0){$+$}}
\put(958,581){\makebox(0,0){$+$}}
\put(961,587){\makebox(0,0){$+$}}
\put(963,592){\makebox(0,0){$+$}}
\put(966,596){\makebox(0,0){$+$}}
\put(968,597){\makebox(0,0){$+$}}
\put(970,598){\makebox(0,0){$+$}}
\put(973,596){\makebox(0,0){$+$}}
\put(975,594){\makebox(0,0){$+$}}
\put(978,590){\makebox(0,0){$+$}}
\put(981,584){\makebox(0,0){$+$}}
\put(985,576){\makebox(0,0){$+$}}
\put(989,567){\makebox(0,0){$+$}}
\put(993,555){\makebox(0,0){$+$}}
\put(999,541){\makebox(0,0){$+$}}
\put(1006,525){\makebox(0,0){$+$}}
\put(1015,505){\makebox(0,0){$+$}}
\put(1029,482){\makebox(0,0){$+$}}
\put(1051,453){\makebox(0,0){$+$}}
\put(1070,435){\makebox(0,0){$+$}}
\put(1070,435){\makebox(0,0){$+$}}
\sbox{\plotpoint}{\rule[-0.500pt]{1.000pt}{1.000pt}}%
\put(1306,632){\makebox(0,0)[r]{$N_{TC}$=9}}
\put(1350,632){\raisebox{-1.2pt}{\makebox(0,0){$\Box$}}}
\put(503,191){\raisebox{-1.2pt}{\makebox(0,0){$\Box$}}}
\put(521,227){\raisebox{-1.2pt}{\makebox(0,0){$\Box$}}}
\put(539,258){\raisebox{-1.2pt}{\makebox(0,0){$\Box$}}}
\put(557,285){\raisebox{-1.2pt}{\makebox(0,0){$\Box$}}}
\put(575,308){\raisebox{-1.2pt}{\makebox(0,0){$\Box$}}}
\put(593,328){\raisebox{-1.2pt}{\makebox(0,0){$\Box$}}}
\put(611,346){\raisebox{-1.2pt}{\makebox(0,0){$\Box$}}}
\put(629,360){\raisebox{-1.2pt}{\makebox(0,0){$\Box$}}}
\put(647,373){\raisebox{-1.2pt}{\makebox(0,0){$\Box$}}}
\put(665,384){\raisebox{-1.2pt}{\makebox(0,0){$\Box$}}}
\put(683,392){\raisebox{-1.2pt}{\makebox(0,0){$\Box$}}}
\put(701,400){\raisebox{-1.2pt}{\makebox(0,0){$\Box$}}}
\put(719,406){\raisebox{-1.2pt}{\makebox(0,0){$\Box$}}}
\put(737,411){\raisebox{-1.2pt}{\makebox(0,0){$\Box$}}}
\put(755,418){\raisebox{-1.2pt}{\makebox(0,0){$\Box$}}}
\put(773,431){\raisebox{-1.2pt}{\makebox(0,0){$\Box$}}}
\put(791,461){\raisebox{-1.2pt}{\makebox(0,0){$\Box$}}}
\put(809,531){\raisebox{-1.2pt}{\makebox(0,0){$\Box$}}}
\put(827,657){\raisebox{-1.2pt}{\makebox(0,0){$\Box$}}}
\put(845,695){\raisebox{-1.2pt}{\makebox(0,0){$\Box$}}}
\put(863,620){\raisebox{-1.2pt}{\makebox(0,0){$\Box$}}}
\put(881,573){\raisebox{-1.2pt}{\makebox(0,0){$\Box$}}}
\put(899,546){\raisebox{-1.2pt}{\makebox(0,0){$\Box$}}}
\put(917,528){\raisebox{-1.2pt}{\makebox(0,0){$\Box$}}}
\put(678,390){\raisebox{-1.2pt}{\makebox(0,0){$\Box$}}}
\put(764,424){\raisebox{-1.2pt}{\makebox(0,0){$\Box$}}}
\put(791,459){\raisebox{-1.2pt}{\makebox(0,0){$\Box$}}}
\put(803,502){\raisebox{-1.2pt}{\makebox(0,0){$\Box$}}}
\put(811,541){\raisebox{-1.2pt}{\makebox(0,0){$\Box$}}}
\put(816,574){\raisebox{-1.2pt}{\makebox(0,0){$\Box$}}}
\put(820,602){\raisebox{-1.2pt}{\makebox(0,0){$\Box$}}}
\put(823,625){\raisebox{-1.2pt}{\makebox(0,0){$\Box$}}}
\put(826,645){\raisebox{-1.2pt}{\makebox(0,0){$\Box$}}}
\put(828,661){\raisebox{-1.2pt}{\makebox(0,0){$\Box$}}}
\put(830,675){\raisebox{-1.2pt}{\makebox(0,0){$\Box$}}}
\put(832,686){\raisebox{-1.2pt}{\makebox(0,0){$\Box$}}}
\put(833,694){\raisebox{-1.2pt}{\makebox(0,0){$\Box$}}}
\put(834,701){\raisebox{-1.2pt}{\makebox(0,0){$\Box$}}}
\put(836,706){\raisebox{-1.2pt}{\makebox(0,0){$\Box$}}}
\put(837,708){\raisebox{-1.2pt}{\makebox(0,0){$\Box$}}}
\put(839,709){\raisebox{-1.2pt}{\makebox(0,0){$\Box$}}}
\put(840,709){\raisebox{-1.2pt}{\makebox(0,0){$\Box$}}}
\put(842,706){\raisebox{-1.2pt}{\makebox(0,0){$\Box$}}}
\put(843,702){\raisebox{-1.2pt}{\makebox(0,0){$\Box$}}}
\put(845,695){\raisebox{-1.2pt}{\makebox(0,0){$\Box$}}}
\put(847,687){\raisebox{-1.2pt}{\makebox(0,0){$\Box$}}}
\put(850,676){\raisebox{-1.2pt}{\makebox(0,0){$\Box$}}}
\put(853,663){\raisebox{-1.2pt}{\makebox(0,0){$\Box$}}}
\put(856,647){\raisebox{-1.2pt}{\makebox(0,0){$\Box$}}}
\put(861,628){\raisebox{-1.2pt}{\makebox(0,0){$\Box$}}}
[6~[6~\put(868,605){\raisebox{-1.2pt}{\makebox(0,0){$\Box$}}}
\put(879,577){\raisebox{-1.2pt}{\makebox(0,0){$\Box$}}}
\put(901,544){\raisebox{-1.2pt}{\makebox(0,0){$\Box$}}}
\put(923,523){\raisebox{-1.2pt}{\makebox(0,0){$\Box$}}}
\end{picture}
\end{center}
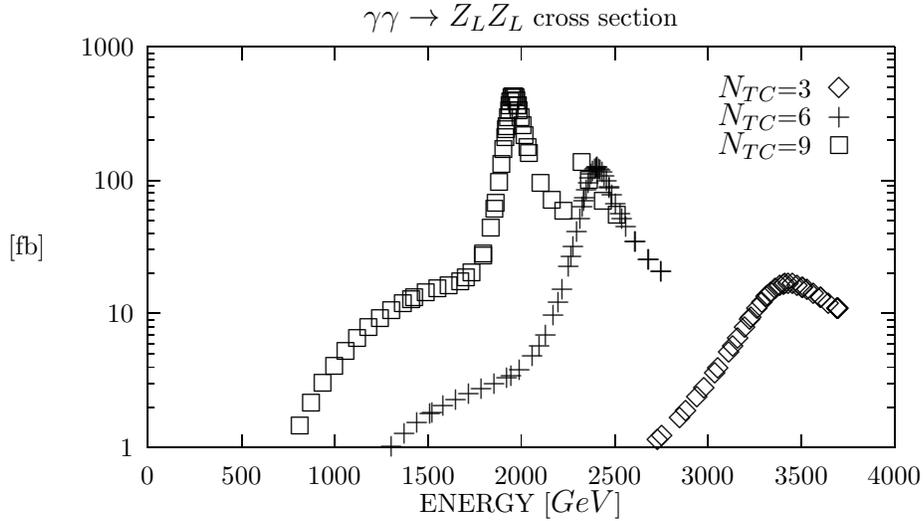
{\bf Figure~10~b.} The total $\gamma\gamma \rightarrow Z_LZ_L$ cross section
as a function of $\sqrt{s}$ and $N_{TC}$. Rescattering and vector meson
dominance effects are similar in magnitude for $\sqrt{s}<(1\, TeV)^2$ but both
are small.

\end{document}